\documentclass[reprint,aps,prb,floatfix,superscriptaddress]{revtex4-2} 

\usepackage{graphicx}
\usepackage{siunitx}
\usepackage{textcomp}
\usepackage{placeins}
\usepackage{gensymb}
\usepackage{amsmath,amssymb}
\usepackage{mathrsfs}
\usepackage{color}
\usepackage[normalem]{ulem}
\usepackage{soul} 
\usepackage{hyperref}
\DeclareUnicodeCharacter{2500}{\textemdash}
\usepackage{placeins}

\newcommand{\gPL}{g_{\smash{\raisebox{-0.4ex}{\scriptsize PL}}}}
\newcommand{\gAbs}{g_{\smash{\raisebox{-0.4ex}{\scriptsize Abs}}}}
\begin{document}

\title{Directional Control of Chiral Polymer Emission Dissymmetry by Plasmonic Metasurfaces}

\author{Debapriya Pal}
\thanks{These authors contributed equally to this work.}
\affiliation{Department of Physics of Information in Matter and
Center for Nanophotonics, NWO-I Institute AMOLF,
Science Park 104, NL 1098 XG Amsterdam, The Netherlands}
\affiliation{Present address: Photonics Initiative,
Advanced Science Research Center, City University of New York,
85 St. Nicholas Terrace, New York, NY 10031, USA}

\author{Louis Minion}
\thanks{These authors contributed equally to this work.}
\affiliation{Department of Materials, Imperial College London,
Prince Consort Road, London SW7 2AZ, United Kingdom}
\affiliation{Department of Chemistry, Molecular Sciences Research Hub,
Imperial College London, London, W12 0BZ, United Kingdom}
\affiliation{Diamond Light Source Ltd, Harwell Science and Innovation Campus, Didcot OX11 0DE, Oxfordshire, United Kingdom}
\affiliation{Present address: Dipartimento di Chimica e Chimica Industriale,
Università di Pisa, Via Moruzzi 13, 56124 Pisa, Italy}

\author{David Reger}
\affiliation{Department of Chemistry, Molecular Sciences Research Hub, Imperial College London, London, W12 0BZ, United Kingdom}
\affiliation{Department of Chemistry, Chemistry Research Laboratory, University of Oxford, 12 Mansfield Road, Oxford, OX1 3TA, United Kingdom}

\author{Jessica Wade}
\thanks{Corresponding author: jessica.wade@imperial.ac.uk}
\affiliation{Department of Materials, Imperial College London, Prince Consort Road, London SW7 2AZ, United Kingdom}

\author{Matthew J. Fuchter}
\thanks{Corresponding author: matthew.fuchter@chem.ox.ac.uk}
\affiliation{Department of Chemistry, Molecular Sciences Research Hub, Imperial College London, London, W12 0BZ, United Kingdom}
\affiliation{Department of Chemistry, Chemistry Research Laboratory, University of Oxford, 12 Mansfield Road, Oxford, OX1 3TA, United Kingdom}

\author{A. Femius Koenderink}
\thanks{Corresponding author: f.koenderink@amolf.nl}
\affiliation{Department of Physics of Information in Matter and Center for Nanophotonics, NWO-I Institute AMOLF, Science Park 104, NL 1098 XG Amsterdam, The Netherlands}

\date{\today}

\begin{abstract}
Circularly polarized (CP) emission is central to next-generation photonic and optoelectronic technologies, yet simultaneously controlling its directionality, intensity, and handedness remains a major challenge. Advances in molecular and materials design have enabled strong chiral light–matter interactions and large dissymmetry between opposite CP states. 
However, conventional thin-film architectures provide limited control over the angular distribution of CP emission and often exhibit strongly angle-dependent chiral emission. Metasurfaces, on the other hand, offer powerful control over emission directionality and polarization, but have predominantly been explored with intrinsically achiral emitters. For the first time, we integrate organic chiral polymer thin films with planar chiral plasmonic metasurfaces to enhance chiral light–matter interactions while controlling the directionality and angular dependence of chiral emission. Using angle-resolved Stokes Fourier polarimetry, we show that metasurface coupling produces directional emission, near-normal dissymmetry factors exceeding 0.7, and, remarkably, direction-dependent reversal of the emission handedness at fixed molecular stereochemistry.
\end{abstract}

\maketitle 
\newpage

\section*{Introduction}
Strong circularly polarized luminescence (CPL) has attracted considerable interest for next-generation optoelectronic technologies~\cite{hubener2021engineering,furlan2024chiral}, including advanced organic light-emitting diodes (OLEDs), high-performance 3-D displays, glare-free screens, quantum technologies, spintronics, and biosensing~\cite{lee2010designs,kim2006potential,hirohata2014future,gisin2002quantum,wagenknecht2010experimental}. Among the various materials investigated for strong CPL, blends of achiral conjugated polymers and small chiral molecules have emerged as particularly promising candidates~\cite{brandt2017added,di2020circularly,deng2021circularly,kim2021chiral,wan2023anomalous,crassous2023materials,chowdhury2025circularly,yang2013induction}. For practical photonic and optoelectronic applications, however, achieving a large dissymmetry factor, which quantifies the intensity imbalance between left- (LH) and right-handed (RH) CP emission, alone is not sufficient. It is also important to control the intensity and degree of CP as a function of emission angle, enabling applications such as directional light routing and polarization-selective detection. Several approaches have been developed to control the directionality and polarization of light emitted by achiral materials, including the use of light-emitting metasurfaces~\cite{vaskin_light-emitting_2019}. Metasurfaces are arrays of subwavelength optical antennas that provide versatile control over the intensity, polarization, and direction of emitted light~\cite{kruk2014spin,raziman2019conditions}. In light-emitting systems, they can tailor emission by coupling to and directing guided optical modes supported within the emissive layer~\cite{lozano2013plasmonics,vaskin_light-emitting_2019}.
Photoluminescence (PL) coupled to laterally propagating modes can acquire spatial coherence before periodic in-plane nanostructures scatter it into selected far-field directions, thereby controlling the angular distribution and polarization of the emitted light. When the in-plane mirror symmetry of the arrangement of meta-atoms within the unit cell is broken, the metasurface can exhibit a chiral optical response. Meinzer~\textit{et al.}~\cite{meinzer2013probing} 
demonstrated that planar-chiral metasurfaces which lack in-plane mirror symmetry, a property commonly referred to as two-dimensional (2D) planar chirality~\cite{plum2009extrinsic}, can generate CP emission from achiral emitters. Subsequent studies have generated and tailored CP emission through in-plane~\cite{cotrufo2016spin,poulikakos2018chiral,gryb2023two,tian2024optimizing,liang2024tailoring} or out-of-plane structural asymmetry~\cite{yin2013interpreting}, while angle-resolved measurements have revealed contributions from both planar chirality and spin–orbit effects~\cite{cotrufo2016spin}. These demonstrations have predominantly involved intrinsically achiral or racemic emitters~\cite{rodrigues2014metamaterials,cotrufo2016spin,vincon2022strong,sun2023enhancing,gryb2023two,ramamurthy2023circularly}. However, the interplay between the intrinsic chirality of an emitter and the structural chirality of a metasurface remains poorly understood, particularly how their combined effects influence the handedness and directionality of CPL.

We explore this interplay experimentally using a chiral polymer platform exhibiting a large emission dissymmetry factor, based on poly(9,9-dioctylfluorene-alt-benzothiadiazole) (F8BT) blended with 10 wt.\% of the enantiopure helicene (aza[6]H) (Figure~\ref{fig:MolecularStructure}a), which combines a strong chiroptical response with solution processability and compatibility with LED architectures~\cite{wan2019inverting,wade2020natural,wan2020highly}. Racemic and enantiopure films can be prepared by selecting the appropriate aza[6]H additive while retaining closely related structural and optoelectronic properties. In films with thicknesses of 140--190~nm, emission dissymmetry factors on the order of 0.1 have been reported~\cite{wan2019inverting}. Despite the relevance of these materials to CP-OLED technologies~\cite{wan2019inverting,wan2020highly}, their full angle-resolved emission dissymmetry remains largely unexplored. Using Fourier-plane Stokes polarimetry~\cite{mohtashami2015angle,rohrich2018quantitative}, we first map the wavevector- and polarization-resolved emission of chiral polymer films on unpatterned substrates, revealing pronounced angular anisotropy. We integrate the films with periodic planar chiral metasurfaces composed of silver nanodimer unit cells that support both localized and lattice resonances. We show that the metasurface geometry can enhance, suppress, or even reverse the emission handedness. These results show how the optical response due to the metasurface can modify the angular distribution and handedness of CP emission from a chiral polymer, even reversing the observed handedness without altering the molecular stereochemistry. More broadly, engineering the spin and momentum of chiral emission independently of molecular chirality through metasurface geometry opens new possibilities for chiral light sources and polarization-selective photonic devices.

\section*{Results}
\subsection*{Absorption and emission properties of chiral films}
\begin{figure}[htbp!]
  \centering
  \includegraphics[width=0.46\textwidth]{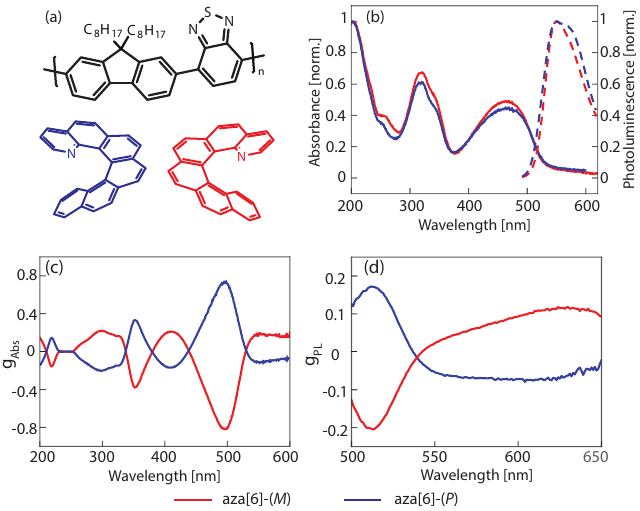}
  \caption{\textbf{Material characterization of chiral blend systems.} (a) Molecular structures of F8BT and the aza[6]H-($P$) (blue) and aza[6]H-($M$) (red) enantiomers. (b) Absorbance (solid) and PL (dashed) spectra of the annealed F8BT:aza[6]H blend films. (c,d) 
  Corresponding absorption and PL dissymmetry spectra, $\gAbs$ and $\gPL$.}
  
  \label{fig:MolecularStructure}
\end{figure}
Figure~\ref{fig:MolecularStructure}a illustrates the molecular structure of F8BT together with the two enantiomers of aza[6]H-($P$) and -($M$). Figure~\ref{fig:MolecularStructure}b shows the absorbance (A) and PL spectra of the corresponding annealed blend films with a thickness ($t$) of around 175 nm. 
As reported previously~\cite{yang2013induction,wan2020highly}, annealed F8BT films blended with enantiomerically pure aza[6]H exhibit large mirror-image circular dichroism (CD) and CPL across the F8BT absorption and emission spectra. In contrast, neat F8BT and F8BT blended with racemic aza[6]H exhibit negligible dissymmetry (Supplementary Fig. S1). The chiroptical contrast is quantified by the dissymmetry factor:
\begin{equation}
g_{\smash{\raisebox{-0.4ex}{\scriptsize X}}} = 2\frac{\text{X}_\text{LH}-\text{X}_\text{RH}}{\text{X}_\text{LH}+\text{X}_\text{RH}},
\end{equation}
where X denotes the absorbance (A) for absorption (Abs) measurements or the emission intensity (I) in case of PL measurements, with subscripts LH and RH denoting left- and right-handed CP, respectively. We refer to these quantities as $\gAbs$ and $\gPL$, shown in Figure~\ref{fig:MolecularStructure}c and d, respectively. 
Importantly, the $\gPL$ spectra are bisignate: the short-wavelength emission has the same handedness as the maximum $\gAbs$ feature, whereas emission near the PL peak ($\sim550$~nm) has the opposite sign, with a maximum $\gPL$ of $\sim0.2$ near 520~nm. Bisignate CPL has also been reported in other chiral molecular and conjugated-polymer systems and has been attributed to a range of mechanisms~\cite{zinna2015lanthanide,duong2017origin,li2023strong}. In F8BT:aza[6]H films, the overall chiroptical response is understood to arise from helicene-induced chiral supramolecular reorganization of the polymer chains~\cite{wade2020natural}.

\subsection*{Angle-resolved polarimetry of unpatterned chiral films}
\begin{figure*}[htbp!]
  \centering
  \includegraphics[width=0.98\textwidth]{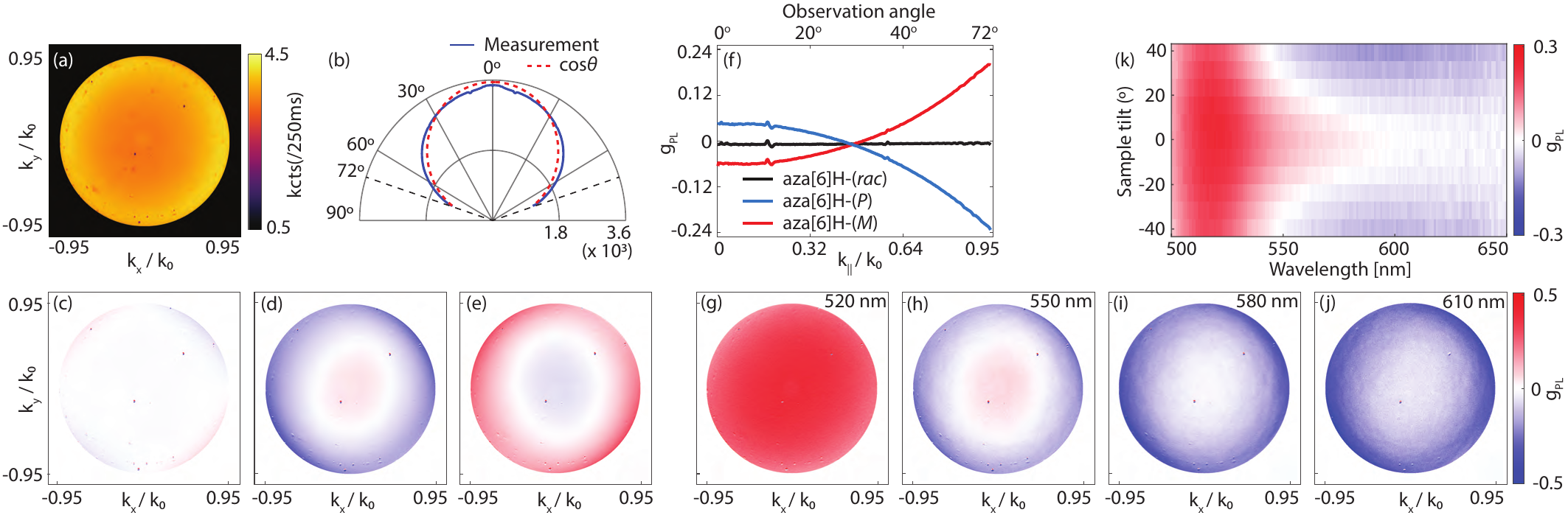}
  \caption{
  \textbf{Angle- and wavelength-resolved chiral emission maps from annealed F8BT:aza[6]H blend films under linearly polarized excitation.}
(a) Panchromatic Fourier-space PL intensity map of an F8BT:aza[6]H-\((rac)\) film.
(b) Corresponding azimuthally averaged emission intensity (blue solid line), compared with a Lambertian \(\cos\theta\) dependence (red dashed line). Black dashed lines indicate the collection limit set by the numerical aperture.
(c--e) Panchromatic Fourier-space \(\gPL\) maps of films containing aza[6]H-\((rac)\), aza[6]H-\((P)\), and aza[6]H-\((M)\), respectively.
(f) Azimuthally averaged \(\gPL\) as a function of normalized in-plane momentum, \(k_{\parallel}/k_0\), and the corresponding observation angle for the -\((rac)\), -\((P)\), and -\((M)\) blend films.
(g--j) Wavelength-resolved Fourier-space \(\gPL\) maps of the F8BT:aza[6]H-\((P)\) film measured using 10 nm full-width at half maximum (FWHM) band-pass filters centered at 520, 550, 580, and 610~nm, respectively.
(k) \(\gPL\) measured using a conventional CPL spectrometer as a function of emission wavelength and sample tilt angle for the F8BT:aza[6]H-\((P)\) film.}
  \label{fig:PlanarPanchromatic}
\end{figure*}

We used back focal plane (Fourier) microscopy to measure the angular distribution of PL from F8BT:aza[6]H blend films under 405 nm excitation. The high numerical aperture (NA $= 0.95$) collection objective provides access to emission up to $72^\circ$ from the surface normal.
Polarimeters in the excitation and detection arms were used to control the pump polarization and perform Stokes polarimetric analysis of the emitted light, enabling evaluation of the angle-resolved $\gPL$. Figure~\ref{fig:PlanarPanchromatic} shows panchromatic angle-resolved emission from annealed F8BT:aza[6]H blend films under linearly polarized excitation, integrated over the full PL band. Figure~\ref{fig:PlanarPanchromatic}a shows the angle-resolved PL intensity from the F8BT:aza[6]H-$(racemic)$ or $(-rac)$ blend. The emission is nearly isotropic in momentum space, with only a modest increase toward the edge of the collection NA, corresponding to larger collection angles. 
The azimuthally averaged radiation pattern in Figure~\ref{fig:PlanarPanchromatic}b is approximately Lambertian, as indicated by the $\cos\theta$ fit~\cite{waynant2000electro}. 

While total intensity patterns are very similar for the racemic and  both enantiopure blends (Supplementary Fig. S2), the corresponding panchromatic $\gPL$ maps reveal qualitatively different behavior. The annealed racemic blend shows negligible CP, with $\gPL \approx 0$ across the Fourier plane (Figure~\ref{fig:PlanarPanchromatic}c), as do the as-cast enantiopure blends (Supplementary Fig.~S3). In contrast, the annealed enantiopure films exhibit a pronounced angular dependence of $\gPL$ (Figure~\ref{fig:PlanarPanchromatic}d,e). The F8BT:aza[6]H-($P$) emission is predominantly LH CP near the normal direction, gradually reversing to RH CP at larger in-plane momenta or wider angles. The F8BT:aza[6]H-($M$) film shows the mirror-imaged response. This trend is summarized in Figure~\ref{fig:PlanarPanchromatic}f, where the azimuthally averaged $\gPL$ remains near zero for the achiral racemic film but evolves with opposite signs for the two chiral, enantiopure blends. 

Because the CPL spectrum is bi-signate (Figure~\ref{fig:MolecularStructure}d), the panchromatic $\gPL$ maps in Figure~\ref{fig:PlanarPanchromatic} average across spectral regions with opposite dissymmetry. To avoid this, we recorded wavelength-resolved angle-dependent $\gPL$ maps for the F8BT:aza[6]H blends using bandpass filters centered at 520, 550, 580, and 610~nm. Figure~\ref{fig:PlanarPanchromatic}g--j shows the wavelength-resolved $\gPL$ maps for F8BT:aza[6]H-($P$) under linear excitation; the complete wavelength- and pump-polarization-resolved datasets for the ($rac$), ($P$), and ($M$) blends are provided in Supplementary Figs.~S4--S6. These data reveal that the emission handedness depends strongly on both wavelength and observation angle. At 520~nm, $\gPL$ is predominantly positive and nearly uniform across the collected momentum space. At 550~nm, the absolute $\gPL$ magnitude is reduced, consistent with the zero crossing of the CPL spectrum (Supplementary Fig. S1b). At longer wavelengths, $\gPL$ becomes predominantly negative, with the strongest dissymmetry emerging at larger emission angles. This trend was independently confirmed by angle-dependent CPL spectroscopy, shown in Figure~\ref{fig:PlanarPanchromatic}k, performed on a rotating stage under depolarized excitation.  Finally, we note that the angular $\gPL$ response also depends weakly on the pump polarization. At the pump wavelength (405 nm), absorption anisotropy is only $\approx$ 1\%~(Figure~\ref{fig:MolecularStructure}c), reflecting a slight preference for one incident handedness over the other.  Fourier-space measurements performed under RH CP, linear, and LH CP excitation (Supplementary Fig.~S7) show modest but systematic pump-dependent changes in the magnitude of $\gPL$, whilst leaving  the dominant wavelength-, angle-, and enantiomer-dependent trends unchanged. This dependence on excitation polarization may be a chiral analogue of the well-known fluorescence polarization memory effect, in which the emitted light retains some of the excitation polarization because the molecular absorption and emission transition dipoles are preferentially aligned.

It is well established that Fresnel coefficients in stratified dielectric stacks can modify the radiation pattern of an emitter. To assess how the stratified environment modifies the emission of chiral dipoles, we modeled archetypical chiral dipoles embedded in a high-index polymer film ($n \approx 2$~\cite{wade2020natural}) sandwiched between semi-infinite air and glass media using a Green's-function formalism (Supplementary section~S4 and Fig.~S8)~\cite{berova2000circular,han2018recent,novotny2012principles,pyramids}. In this model, the electric and magnetic dipoles are coupled with a quarter-wave phase shift, with the sign of the phase encoding the handedness. The simulations reproduce the reversal of emission handedness between opposite enantiomers and an approximately Lambertian far-field intensity distribution into air. However, the simulated $\gPL$ maps exhibit only weak angular variation, in stark contrast to the strong angular dependence observed experimentally in Figure~\ref{fig:PlanarPanchromatic}d--j. This discrepancy suggests that far-field photonic effects alone cannot account for the observed behavior. Possible contributing factors include: limitations of the chiral point dipole model, since emission in these films is thought to arise from excitons delocalized along twisted polymer chains and coupled across neighboring chains~\cite{laidlaw2021factors}, handedness-selective filtering of waveguided modes by CD, given the spectral overlap of CD and PL prior to outcoupling, and polarization rotation during propagation due to combined twisting linear and circular birefringence. Previous morphology studies with soft X-ray scattering at the carbon K-edge have revealed in-plane periodicities of 260--330 nm in these films~\cite{wade2020natural}, which could provide the necessary scattering contrast for emission outcoupling after propagation in the waveguide. Further work is needed to establish how molecular orientation, higher-order optical transitions, and reabsorption processes contribute to the observed angular dependence of the emission dissymmetry factor.

\subsection*{Reshaping chiral emission using metasurfaces}
\begin{figure}[htbp!]
  \centering
  \includegraphics[width=0.48\textwidth]{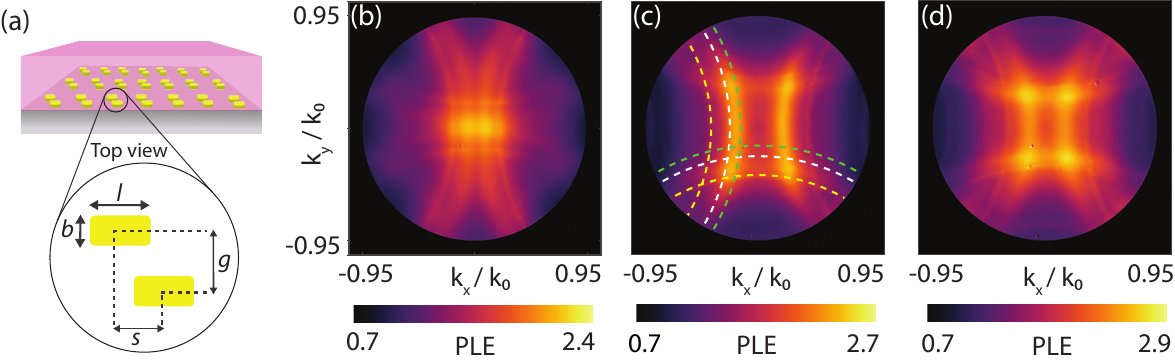}
  \caption{\textbf{Structural tuning of diffractive PL outcoupling.}
(a) Schematic of the metasurface array on glass (gray), coated with the emissive F8BT:aza[6]H film (pink). Inset: a unit cell comprising two rectangular nanorods (yellow) with length \(L=75\) nm, width \(b=50\) nm, height \(h=35\) nm and center-to-center separation $g=100$~nm.
(b--d) Fourier PLE maps of annealed F8BT:aza[6]H-\((P)\) films on achiral metasurfaces (\(s=0\)), measured using 10-nm-FWHM band-pass filters.
(b) \(325\) nm pitch at \(550\) nm.
(c) \(325\) nm pitch at \(610\) nm, with calculated momentum-matching contours for TE modes (green, \(n_{\mathrm{TE}}=1.73\)), TM modes (white, \(n_{\mathrm{TM}}=1.63\)) and diffraction into glass (yellow, \(n_{\mathrm{glass}}=1.47\)), referenced to the \((-1,0)\) and \((0,-1)\) reciprocal lattice vectors.
(d) \(340\) nm pitch at \(610\) nm.
}
  \label{fig:ExpFourier}
\end{figure}

We next consider how the angular distribution and handedness of intrinsically chiral emission can be actively reshaped using metasurfaces. Here, we employ planar plasmonic metasurfaces to engineer the CP emission from intrinsically chiral emitters. Building on the approaches introduced by Meinzer \textit{et al.}~\cite{meinzer2013probing} and Cotrufo \textit{et al.}~\cite{cotrufo2016spin}, we designed and fabricated metasurfaces tailored to F8BT:aza[6]H blends using e-beam lithography. The design has unit cells formed by two parallel silver nanorods separated by \(g=100\)~nm along their short axes (Figure~\ref{fig:ExpFourier}a) placed on a square grid of pitch $a= 325$~nm, chosen to match the polymer emission maximum ($\lambda = 550$ nm) to the polymer layer waveguide mode ($a = \lambda/n_{\text{WG}}$ with  waveguide mode index $n_\text{WG} \approx 1.69$~\cite{urbach1998spontaneous}). This condition enables Bragg outcoupling of waveguided emission toward the sample normal. When the rods are perfectly aligned ($s=0$), the unit cell retains in-plane mirror symmetry, and its normal-incidence transmission, reflection, and absorption responses are achiral. Planar chirality is introduced by shifting the rods in opposite directions along their long axes ($s \neq 0$), which breaks the in-plane mirror symmetry. We estimate the polarization-resolved PL enhancement (PLE), defined as the metasurface-assisted emission normalized to that of the unpatterned film, using the Helmholtz reciprocity principle implemented in finite-element simulations in COMSOL for an achiral polymer layer (Supplementary section~S5 and Supplementary Fig.~S9). The simulated spectra qualitatively reproduce the response reported previously for related planar chiral metasurfaces in achiral environments~\cite{meinzer2013probing,cotrufo2016spin}.
The polymer layer supports TE- and TM-like guided modes that give rise to distinct lattice resonances near the emission band, while the localized plasmon resonance lies at longer wavelengths. For the achiral metasurface ($s=0$), LH and RH CP emission are enhanced equally, yielding $\gPL=0$. Introducing planar chirality ($s\neq0$) produces a finite dissymmetry, reaching $|\gPL|\approx0.15$ (Supplementary Fig.~S9b), with the stronger response associated with the TM-like resonance. This enhanced  sensitivity for the TM mode likely reflects its longitudinal and out-of-plane field components, which interact more strongly with the asymmetric dielectric environment of the nanostructure and layer interfaces, thereby enhancing the chiral response.

\begin{figure*}[htbp!]
  \centering
  \includegraphics[width=\textwidth]{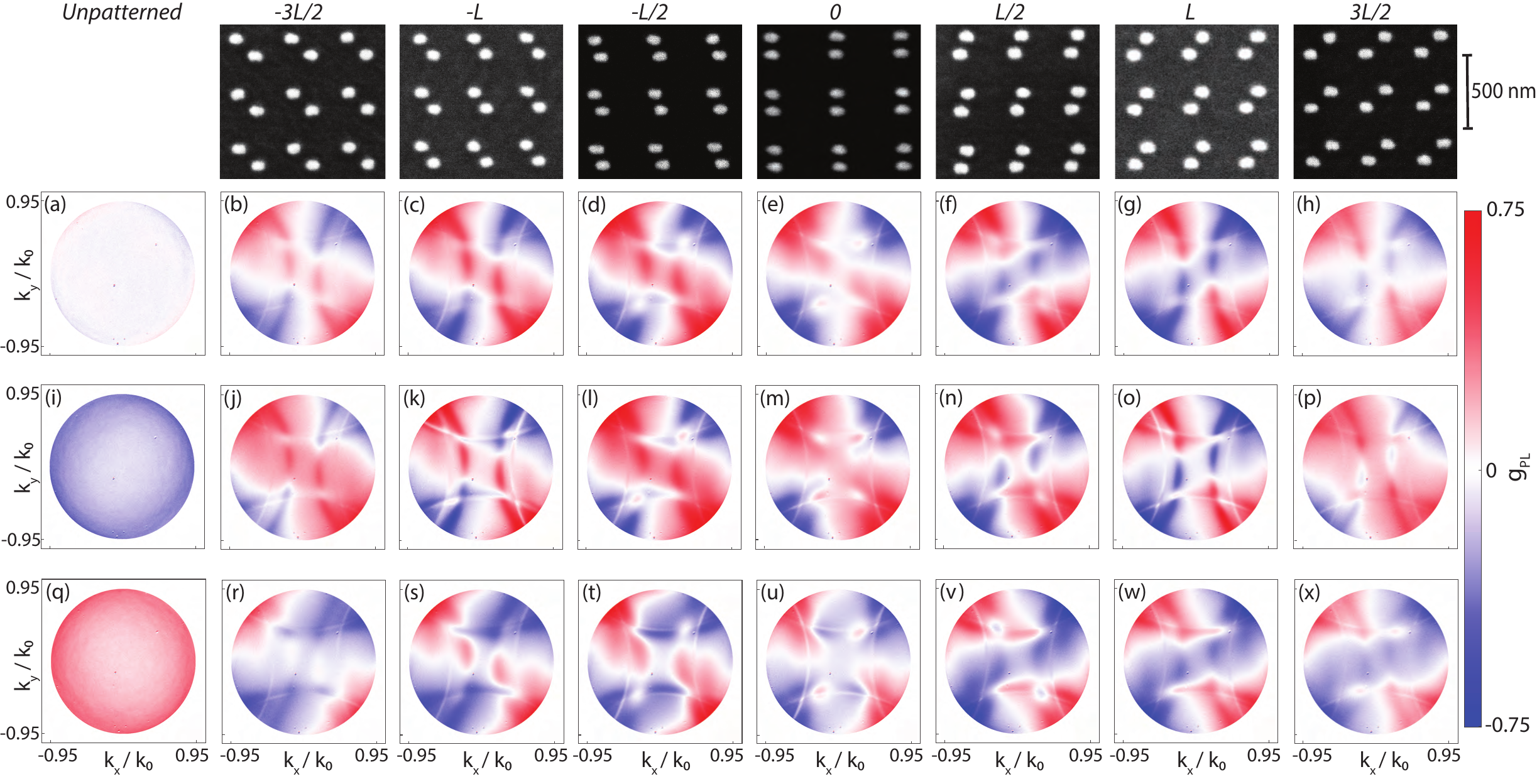}
  \caption{\textbf{Metasurface-controlled angle-resolved chiral emission from F8BT:aza[6]H blend films.}
Top: SEM of the fabricated metasurfaces with a pitch of 325~nm and nanorod shifts \(s=-3L/2,-L,-L/2,0,L/2,L,\) and \(3L/2\). Scale bar, 500~nm. Variations in image contrast arise from charging and carbon deposition on the Elektra layer.
(a--h) Fourier-space \(\gPL\) maps at 610~nm for F8BT:aza[6]H-\((rac)\) films under linearly polarized excitation: (a) unpatterned film reference and (b--h) metasurfaces with the shifts indicated above.
(i--p) Corresponding maps for F8BT:aza[6]H-\((P)\) films under LH CP excitation.
(q--x) Corresponding maps for F8BT:aza[6]H-\((M)\) films under RH CP excitation.}
  \label{fig:chiralmetasurface}
\end{figure*}

Before introducing planar chirality, we first establish how the achiral lattice controls the angular intensity distribution of chiral polymer emission. The measured Fourier-plane PLE maps from F8BT:aza[6]H-\((P)\) films show a diffuse emission background together with pronounced directional enhancements whose positions depend on both wavelength and lattice pitch (Figure~\ref{fig:ExpFourier}b--d). Although the emission originates from spatially incoherent emitters, a substantial fraction couples to laterally propagating guided modes supported by the polymer film. This provides the spatial coherence required for the periodic lattice to Bragg-scatter the emission into well-defined far-field directions, producing sharp diffraction features along the waveguide-mode contours~\cite{bailly2021spatial,pal2024metasurface}.  For a 325~nm pitch, the 550~nm map shows enhanced emission near the sample normal, whereas at 610~nm the enhanced features move to larger in-plane momenta. Increasing the pitch to 340~nm (Figure~\ref{fig:ExpFourier}d) shifts these features back toward the center of the light cone, consistent with the expected Bragg condition. The feature positions retain the four-fold symmetry of the square lattice, while their intensities are anisotropic, with stronger emission along the vertical diffraction direction due to the nanodimer unit-cell geometry, which preferentially couples to $x$-polarized dipolar modes of the nanorods.

\subsection*{Chirality of the metasurface modified emission}
We next investigate the polarization properties of the emitted light. First, we consider the angular \(\gPL\) maps of the achiral F8BT:aza[6]H-($rac$) blend film on unpatterned substrate and on metasurfaces with different lateral shifts \(s\), with emission centered at 610~nm under linear pump polarization (Figure~\ref{fig:chiralmetasurface}a--h). The unpatterned film shows negligible \(\gPL\) across the Fourier plane, as expected for an achiral emitter ensemble (Figure~\ref{fig:chiralmetasurface}a). In contrast, coupling the same racemic film to the metasurface produces large angle-dependent \(\gPL\). For the achiral metasurface \((s=0)\), the \(\gPL\) map (Figure~\ref{fig:chiralmetasurface}e) exhibits an antisymmetric pattern with sign reversal across both the \(k_x\) and \(k_y\) axes. This response arises from the chiral experimental configuration formed by the incident wavevector, sample geometry, and oblique collection direction, even though both the emitter and metasurface are achiral. Similar chiroptical responses from achiral structures under oblique observation have been described as extrinsic chirality or pseudo-chirality~\cite{plum2009extrinsic,sersic2012ubiquity,lindell1994electromagnetic}, with spin--orbit coupling providing the underlying mechanism~\cite{bliokh2015spin}. The antisymmetric angular response is consistent with symmetry constraints: when integrated over a symmetry-complete back aperture, the net chirality vanishes, even though strongly dissymmetric emission can appear in specific directions~\cite{poulikakos2016optical}.

We then examine how intrinsically chiral emission is reshaped by chiral metasurfaces (Figure~\ref{fig:chiralmetasurface}i--x and Supplementary Figs.~S10--S12). Relative to the unpatterned film references, coupling to the metasurfaces strongly reshapes the intrinsic angular \(\gPL\) response of both enantiopure blends. For F8BT:aza[6]H-\((P)\), the unpatterned film emits predominantly RH CPL at 610~nm, whereas the metasurface can bias the emission toward LH CPL, reversing the dominant handedness. This handedness reversal due to the metasurface can be rationalized by considering two emission pathways in the F8BT:aza[6]H film-on-glass geometry: direct emission into the far field and guided emission that propagates laterally within the high-index polymer film before metasurface-assisted outcoupling. Because of the high refractive index of F8BT:aza[6]H, we estimate that only $\sim15\%$ of the emission escapes directly to the far field; the metasurface-enhanced signal is therefore expected to be dominated by the outcoupled guided component~\cite{pal2024metasurface}. Using an absorptance \(A \approx 0.05\) for a film thickness \(t=175~\mathrm{nm}\), the absorption length is \(L_\mathrm{abs}=-t/\ln(1-A)\approx3.5~\mu\mathrm{m}\), nearly six times larger than the selected emission wavelength of 610~nm. This estimated propagation length is sufficient for reabsorption and polarization evolution to plausibly influence the handedness of the observed outcoupled emission. Since \(\gAbs\) and \(\gPL\) at \(\lambda=610~\mathrm{nm}\) have the same sign (Figure~\ref{fig:MolecularStructure}c,d), handedness-selective reabsorption can preferentially attenuate the intrinsically favored CP component during guided propagation and contribute to opposite handedness emission after metasurface-assisted outcoupling. Thus, the guided emission is polarization-filtered during propagation before it is outcoupled by the metasurface. As in the racemic film coupled to the achiral \((s=0)\) metasurface (Figure~\ref{fig:chiralmetasurface}e), opposite handedness appears in diagonally opposite quadrants at large angles. For the enantiopure films, however, this quadrant pattern is superimposed on an overall handedness bias from the chiral emitter, leading to a non-zero integrated \(\gPL\). For example, for the F8BT:aza[6]H-\((P)\) blend on the \(s=0\) metasurface, LH CP emission dominates overall, corresponding to a net positive \(\gPL\) (Figure~\ref{fig:chiralmetasurface}m). Inverting the lateral shift \(s\) of the nanodimer unit cell reverses the 
induced handedness of the vertically oriented diffraction features associated with waveguide-mode outcoupling, with the strongest contrast near \(|s|=L\). The central outcoupled SLR features retain a slight bias toward LH CP emission, consistent with the intrinsic chiral contribution from the F8BT:aza[6]H-\((P)\) blend. This sign reversal demonstrates that the emitted handedness is not determined solely by molecular chirality, but can be selected by the planar chiral metasurface geometry.

Finally, comparing enantiopure chiral polymer blends on metasurfaces with opposite geometrical shifts, \(\pm s\), provides insight into how intrinsic molecular chirality, pump handedness, and metasurface geometry jointly determine the handedness of CPL. For metasurface-coupled films, the pertinent symmetry relation in the experiment is that identical angular \(\gPL\) maps up to a sign change are obtained when reversing \emph{both} the molecular handedness (F8BT:aza[6]H-\((P)\) versus -\((M)\) films) \emph{and} the metasurface handedness ($+$ to $-$ $s$) at the same time.  This relation is confirmed experimentally by comparing Fig.~\ref{fig:chiralmetasurface}p for the F8BT:aza[6]H-($P$) film at the largest positive meta-atom shift ($s=3L/2$) with Fig.~\ref{fig:chiralmetasurface}r for the F8BT:aza[6]H-($M$) film at the corresponding negative shift ($s=-3L/2$).  In these experiments, we took care to also flip the pump polarization handedness, to account for the weak intrinsic pump polarization.  Together, these results show that angular CP emission is governed by the interplay of intrinsic molecular chirality, guided-mode propagation, and planar chiral metasurface geometry, with pump handedness providing an additional degree of control.

\section*{Discussion}
We demonstrate that planar chiral metasurfaces provide an effective strategy for controlling CPL from intrinsically chiral emissive thin films. Fourier-plane polarimetry reveals that unpatterned chiral polymer films exhibit large CPL, with a \(\gPL\) that depends strongly on wavelength and observation angle, and modestly on pump polarization. This angular dependence underscores the importance of reporting the collection numerical aperture in spectroscopic measurements of chiral optoelectronic systems, since angular averaging can obscure signatures of intrinsic chirality. Angle-resolved $k$-space polarimetry offers a useful diagnostic for probing chiral light-matter interactions and evaluating CP-(O)LED architectures, where scrambling of CP at reflective electrodes remains a key bottleneck~\cite{zinna_design_2017}.
By coupling these films to laterally shifted nanorod metasurfaces, we show that the angular distribution of \(\gPL\) can be reshaped through diffractive outcoupling of guided modes. Whereas unpatterned chiral polymers exhibit their largest \(\gPL\) mainly at high emission angles, the metasurfaces provide tunable angular control over the emission handedness and enhance the near-normal dissymmetry to values approaching $|\gPL|\approx0.7$. The dominant handedness of metasurface-outcoupled emission is reversed relative to the same thin film on an unpatterned substrate, which we attribute to selective extraction of waveguided emission that undergoes handedness-specific reabsorption during propagation. Planar metasurfaces therefore offer a promising route toward polarization-engineered chiral light sources with enhanced directionality and tunability. More broadly, integrating intrinsically chiral emitters with nonlocal metasurfaces~\cite{overvig2021thermal} could extend chiral emission control beyond handedness and directionality toward shaping the spatial coherence and vectorial wavefront of incoherent light, opening routes to structured chiral emission and vectorial holography directly at the source.

\section*{Methods}
\subsection*{Materials and thin film preparation}
Aza[6]H enantiomers and F8BT polymer were synthesized, purified, and separated following standard protocols~\cite{weimar2013scalable, troughton2020universal}, with synthesis details provided in Supplementary section~S1. F8BT and aza[6]H were dissolved in toluene (35~mg/mL) and blended to prepare 10~wt\% aza[6]H:F8BT solutions for both enantiomers and the racemic mixture. Glass substrates were cleaned by sequential sonication in acetone and isopropyl alcohol (IPA), followed by 5 minutes of O$_2$ plasma treatment. Films were fabricated by spin-coating the blend solution at 2500~rpm for 60~seconds, yielding a thickness of approximately 175~nm, as measured by a KLA-Tencor optical profilometer. The spin-coated films were then thermally annealed at $140^\circ$C for 10~minutes on a hotplate under ambient conditions inside a fume hood.

\subsection*{Optical characterization}
\subsubsection*{CD and CPL}
CD, absorption, and CPL measurements shown in Figure~\ref{fig:MolecularStructure} were performed on similarly prepared films. CD and absorption spectra were acquired using a Chirascan V100 CD spectrometer. CPL measurements were conducted using a custom-built spectrometer described previously~\cite{moreno2024enhancing}, with excitation provided by a depolarized 405 nm LED. Emission from the sample was first collimated and passed through a photoelastic modulator (PEM-100, Hinds Instruments), followed by a Glan--Thompson polarizer (Thorlabs) oriented at $45^\circ$ with respect to the PEM axis, then focused into a monochromator (Bentham Instruments) and detected with a PMT (Hamamatsu H13226A-103). The photocurrent was amplified (Femto) and demodulated by a lock-in amplifier (Signaloc 2100). The measured in-phase lock-in signal, corrected for the calibrated phase offset and denoted as $X_{\mathrm{corr}}$, was used to calculate the emission dissymmetry factor as
\begin{equation}
\gPL =
\frac{\sqrt{2}\,X_{\mathrm{corr}}}
{J_1(\pi/2)\,DC},
\end{equation}
where $J_1$ is the first-order Bessel function and $DC$ is the time-averaged detector signal~\cite{kitzmannfundamentals2023}. For angle-dependent spectra, samples were mounted onto a micrometer-adjustable rotating mount with an iris to isolate the central emission. A 405~nm LED was depolarized with a quartz depolarizer and directed onto the sample after reflection from a 425~nm long pass dichroic mirror. A setup schematic, along with the setup details, is provided in Supplementary Fig.~S13.

\subsubsection*{Fourier microscope}
Fourier polarimetry was performed on a custom inverted microscope using a 405~nm CW laser (Cobolt 06-MLD) for excitation and a Zeiss N-Achroplan 63X, NA~$=0.95$ objective (no coverglass) for both illumination and collection from the air side~\cite{pal2025dark}. A polarimeter (polarizer and quarter-wave plate) controlled the pump polarization. Emission was spectrally filtered using a dichroic long-pass filter and imaged onto a CCD through a 200~mm tube lens. Polarization-resolved angular patterns were acquired by inserting a Fourier lens at the objective's back focal plane. A second polarimeter in the detection path enabled full polarization analysis, and both average intensity and $\gPL$ maps were reconstructed from polarimetric images. A full schematic is provided in Supplementary Fig.~S14.

\subsection*{Numerical simulation}
The optical dipole simulations for archetypal chiral emitters were performed using the Green's function formalism. The electric field at an observation point $\mathbf{r}$ due to a point dipole source located at $\mathbf{r}_0$ was calculated via the free-space electric Green's tensor $G(\mathbf{r}, \mathbf{r}_0)$~\cite{novotny2012principles}. To account for the layered environment, $k_\parallel$-dependent Fresnel reflection coefficients were incorporated for both $s$- and $p$-polarized components. The angular radiation pattern was obtained by integrating the time-averaged Poynting vector of the asymptotic far-field electric field, yielding the differential radiated power per unit solid angle. The optical response of the metasurface in the presence of an emitter layer was simulated using the finite element method (FEM) in COMSOL Multiphysics, employing full three-dimensional models under plane-wave excitation. The computational domain included top and bottom ports to launch and receive the electromagnetic fields, and was terminated with perfectly matched layers (PMLs) along the vertical direction to suppress reflections. Bloch-Floquet periodic boundary conditions were applied in the lateral directions to model infinite metasurface periodicity. Further implementation details are provided in Supplementary sections~S4 and S5.

\subsection*{Metasurface Fabrication}
Glass substrates were cleaned by sequential sonication, base piranha treatment, and O$_2$ plasma. A $\sim$150~nm PMMA 495-A8 layer was spin-coated, followed by thermal evaporation of a $\sim$20~nm Germanium (Ge) hard mask at 0.05~nm/s using a Polyteknik E-Flex evaporator. Next, $\sim$60~nm of CSAR AR-P 6200:09 resist was spin-coated and patterned via e-beam lithography (Raith Voyager, 50~keV) at a nominal dose of 130~$\mu$C/cm$^2$. After development (Pentylacetate, O-Xylene, IPA), the Ge and PMMA layers were etched using a mixture of SF$_6$ and O$_2$ plasma. A 35~nm silver (Ag) layer was evaporated at a rate of 0.1~nm/s, followed by a lift-off in warm acetone. The resulting metasurface arrays have lateral footprints of 150~$\mu$m~$\times$~150~$\mu$m each. SEM images were captured after spinning a thin 60~nm Elektra layer on top of the structures, which was rinsed twice with water prior to deposition of the polymer layer.

\section*{Supplementary information}
Additional details on material synthesis and chiroptical characterization, extended angle-resolved polarimetry of unpatterned and metasurface-coupled films, numerical modelling of chiral emission and metasurface optical response, and experimental methods and setup schematics.

\section*{Acknowledgements and funding}
The authors thank Dr. Hussain Al-Tuwailib for preparing one of the F8BT polymer batches used in this study. The work of D.P. and A.F.K. is part of the Dutch Research Council (NWO). The work of D.P. is part of the NanoLED project (project number 17100) under the research programme High Tech Systems and Materials 2018, which is partly financed by NWO. D.P. and A.F.K. acknowledge financial support from the European Commission under the EIC Pathfinder
CHIRALFORCE project (Grant Agreement No. 101046961). The authors acknowledge support from the Centre for Doctoral Training in Advanced Characterization of Materials (Grant Ref: EP/S023259/1). This material was based upon work supported by the Air Force Office of Scientific Research under award number FA9550-23-1-0633. L.M. was additionally supported by a PhD studentship from the Diamond Light Source (DLS). J.W. thanks the Royal Society (URF/R1/23173). 

\section*{Declarations}
\subsection*{Author contributions}
D.P., A.F.K., M.J.F., and J.W. conceived the idea. D.P. performed film preparation, metasurface fabrication, Fourier measurements, simulations, and data analysis. L.M. performed variable angle CPL spectroscopy measurements and data analysis. D.R. performed material synthesis and characterization. A.F.K., M.J.F., and J.W. supervised the work. D.P. and A.F.K. wrote the manuscript with feedback from all authors. All authors reviewed and finalized the manuscript.

\subsection*{Data availability}
The data supporting the findings of this study are available within the article and its Supplementary information. Additional data are available from the corresponding authors upon reasonable request.

\subsection*{Competing interests}
The authors declare no competing interests.

\bibliography{references.bib}
\end{document}


\title{Supplementary Information: Directional Control of Chiral Polymer Emission Dissymmetry by Plasmonic Metasurfaces}

\author{Debapriya Pal}
\thanks{These authors contributed equally to this work.}
\affiliation{Department of Physics of Information in Matter and
Center for Nanophotonics, NWO-I Institute AMOLF,
Science Park 104, NL 1098 XG Amsterdam, The Netherlands}
\affiliation{Present address: Photonics Initiative,
Advanced Science Research Center, City University of New York,
85 St. Nicholas Terrace, New York, NY 10031, USA}

\author{Louis Minion}
\thanks{These authors contributed equally to this work.}
\affiliation{Department of Materials, Imperial College London,
Prince Consort Road, London SW7 2AZ, United Kingdom}
\affiliation{Department of Chemistry, Molecular Sciences Research Hub,
Imperial College London, London, W12 0BZ, United Kingdom}
\affiliation{Diamond Light Source Ltd, Harwell Science and Innovation Campus, Didcot OX11 0DE, Oxfordshire, United Kingdom}
\affiliation{Present address: Dipartimento di Chimica e Chimica Industriale,
Università di Pisa, Via Moruzzi 13, 56124 Pisa, Italy}

\author{David Reger}
\affiliation{Department of Chemistry, Molecular Sciences Research Hub, Imperial College London, London, W12 0BZ, United Kingdom}
\affiliation{Department of Chemistry, Chemistry Research Laboratory, University of Oxford, 12 Mansfield Road, Oxford, OX1 3TA, United Kingdom}

\author{Jessica Wade}
\thanks{Corresponding author: jessica.wade@imperial.ac.uk}
\affiliation{Department of Materials, Imperial College London, Prince Consort Road, London SW7 2AZ, United Kingdom}

\author{Matthew J. Fuchter}
\thanks{Corresponding author: matthew.fuchter@chem.ox.ac.uk}
\affiliation{Department of Chemistry, Molecular Sciences Research Hub, Imperial College London, London, W12 0BZ, United Kingdom}
\affiliation{Department of Chemistry, Chemistry Research Laboratory, University of Oxford, 12 Mansfield Road, Oxford, OX1 3TA, United Kingdom}

\author{A. Femius Koenderink}
\thanks{Corresponding author: f.koenderink@amolf.nl}
\affiliation{Department of Physics of Information in Matter and Center for Nanophotonics, NWO-I Institute AMOLF, Science Park 104, NL 1098 XG Amsterdam, The Netherlands}

\maketitle
\newpage
\tableofcontents
\newpage
\section{Synthesis of chemicals}
\subsection{Poly(9,9-dioctylfluorene-alt-benzothiadiazole) (F8BT)}
The synthetic procedures for F8BT and its precursor 2,2'-(9,9-dioctyl-9H-fluorene-2,7-diyl)bis(4,4,5,5-tetramethyl-1,3,2-dioxaborolane) were adapted from the literature~\cite{troughton2020universal}. The precursors 4,7-dibromo-2,1,3-benzothiadiazole and 2,7-dibromofluorene were purchased from TCI. Before polymerization, all precursors were purified by small-molecule recycling GPC.
\begin{figure*}[ht]
  \centering
  \includegraphics[width=0.98\textwidth]{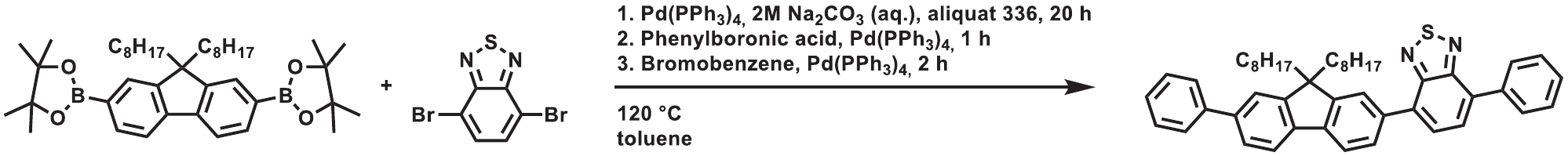}
\end{figure*}
Under a nitrogen atmosphere, a 2-neck round-bottom flask (50 mL), equipped with a magnetic stirring bar and a reflux condenser, was charged with 4,7-dibromobenzo[c]-1,2,5-thiadiazole (200 mg, 0.680 mmol), 2,2'-(9,9-dioctyl-9H-fluorene-2,7-diyl)bis(4,4,5,5-tetramethyl-1,3,2-dioxaborolane) (446 mg, 0.694 mmol), tetrakis(triphenylphosphine)palladium (15.7 mg, 13.6 $\mu$mol), aliquot 336 (one drop) and toluene (7.5 mL). The reaction mixture was degassed via nitrogen bubbling and finally degassed, 2M aq. Na2CO3 (2.5~mL) was added. The reaction mixture was stirred for 20~hr at 120$^o$~C. A degassed solution of phenylboronic acid (82.9 mg, 0.68 mmol) in toluene (5 mL), together with tetrakis(triphenylphosphine)palladium (15.7 mg, 13.6 $\mu$mol), was added, and the reaction mixture was stirred for 1~hr at 120$^o$~C. A solution of bromobenzene (71.0 $\mu$L, 0.68 mmol) in toluene (5 mL) together with tetrakis(triphenylphosphine)palladium (15.7 mg, 13.6 $\mu$mol) was added, and the reaction mixture was stirred for another 2 h at 120$^o$~C. The mixture was allowed to cool to room temperature, and the phases were separated. The aqueous phase was extracted with CHCl3 (50 mL). The combined organic phases were dried over MgSO4, filtered off, and the solvent was concentrated to ~10 mL. The concentrated solution was added dropwise to cold, stirred MeOH (150 mL) to precipitate the crude product. The solid was collected in a cellulose thimble and purified with Soxhlet extractions with methanol, acetone, and hexane respectively. The hexane fraction was collected and concentrated to ~10 mL. The concentrated solution was added dropwise to cold, stirred MeOH (150~mL) to precipitate the polymer. After filtration and drying under vacuum, the polymer was obtained as a dark orange solid with a yield of 103 mg.
GPC analysis: M$_\text{n} =$ 10k, M$_\text{w} =$ 19k, PD $=$ 1.98.

F8BT used at Imperial College was provided by Cambridge Display Technology (UK, company number: 02672530) and used, as received, while the F8BT used at AMOLF was synthesized at Imperial College following the procedure mentioned above.
\subsection{Aza[6]helicene}
Aza[6]H was synthesized following previously reported procedures~\cite{weimar2013scalable}. Chiral resolution was carried out by Reach Separation, UK, via chiral HPLC. To remove other small impurities, the individual enantiomers were additionally purified by small-molecule GPC before use. 

\section{Chiroptical characterization}
The CD and CPL spectra of the annealed enantiopure blend films are shown in Fig.~S1. The aza[6]H-($P$) and aza[6]H-($M$) films exhibit approximately mirror-symmetric chiroptical responses, confirming the opposite handedness of the two enantiomeric blends. These measurements were used to calculate the dissymmetry spectra presented in Fig.~1 of the main text.

\begin{figure*}[ht]
  \centering
  \includegraphics[width=0.95\textwidth]{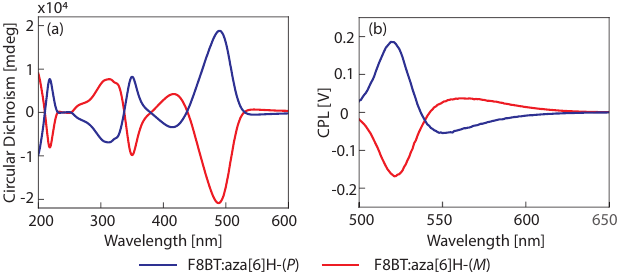}
  \caption{\textbf{Material characterization of chiral blend films.} (a) CD spectra and (b) CPL spectra of annealed F8BT:aza[6]H films containing aza[6]H-($P$) (blue) and aza[6]H-($M$) (red).}
  \label{fig:MolecularStructure}
\end{figure*}

\section{Extended angle-resolved measurements of unpatterned films}
To assess the influence of thermal annealing and pump polarization on the angular chiral emission, we measured Fourier-space $\gPL$ maps for racemic and enantiopure F8BT:aza[6]H blend films. 

\begin{figure*}[ht]
  \centering
  \includegraphics[width=0.84\textwidth]{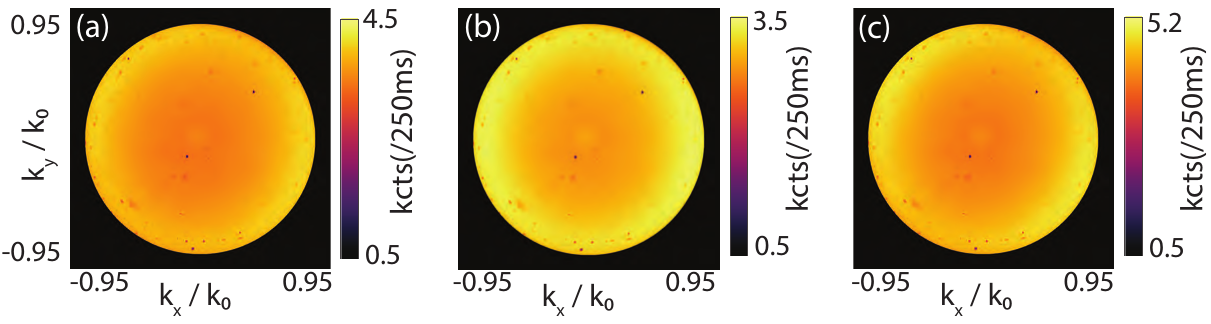}
  \caption{\textbf{Panchromatic PL intensity maps from annealed F8BT:aza[6]H blend films.} Fourier-space intensity maps under linear polarized excitation containing (a) aza[6]H-($rac$), (b) aza[6]H-($P$), and (c) aza[6]H-($M$).}
  \label{fig:Pan_intensity_afterAnnealing}
\end{figure*}

\vspace{1cm}

Figure~S2 shows that the total panchromatic intensity distributions are nearly identical for the racemic and both enantiopure blends. The corresponding intensity distribution for F8BT:aza[6]H-($rac$), shown in Fig.~S2a, is also presented in Fig.~2a of the main text.
\vspace{1cm}

Before annealing, all three blends exhibit comparatively weak angle-resolved chiral dissymmetry, as shown in Fig.~S3. The pronounced angular response discussed in the main text emerges primarily after thermal annealing.
\\\\\\\\\

\begin{figure*}[ht]
  \centering
  \includegraphics[width=0.75\textwidth]{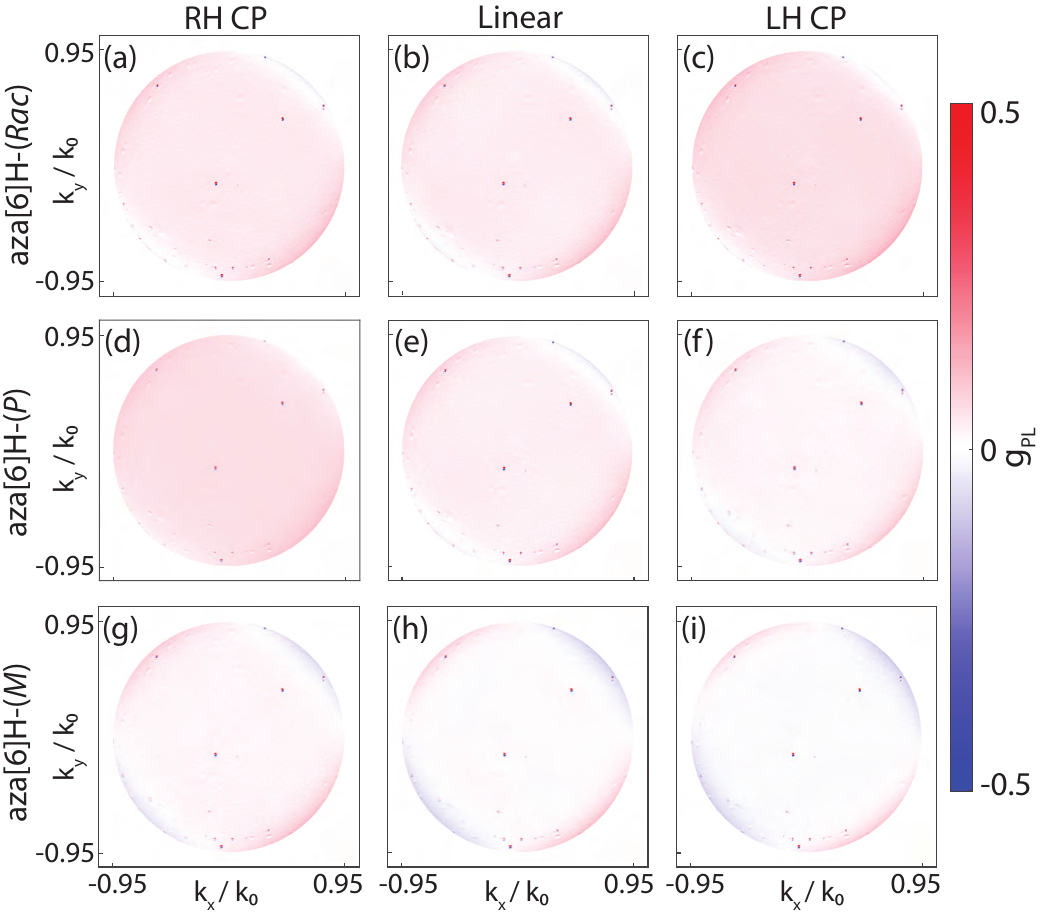}
  \caption{\textbf{Panchromatic angle-resolved $\gPL$ maps from as-cast/ before annealing F8BT:aza[6]H blend films.} Fourier-space $\gPL$ maps measured under three pump polarizations: RH CP, linear, and LH CP excitation (columns). Rows correspond to films containing (a--c) aza[6]H-($rac$), (d--f) aza[6]H-($P$), and (g--i) aza[6]H-($M$).}
  \label{fig:Pan_BeforeAnnealing}
\end{figure*}

\begin{figure*}[ht]
  \centering
  \includegraphics[width=0.9\textwidth]{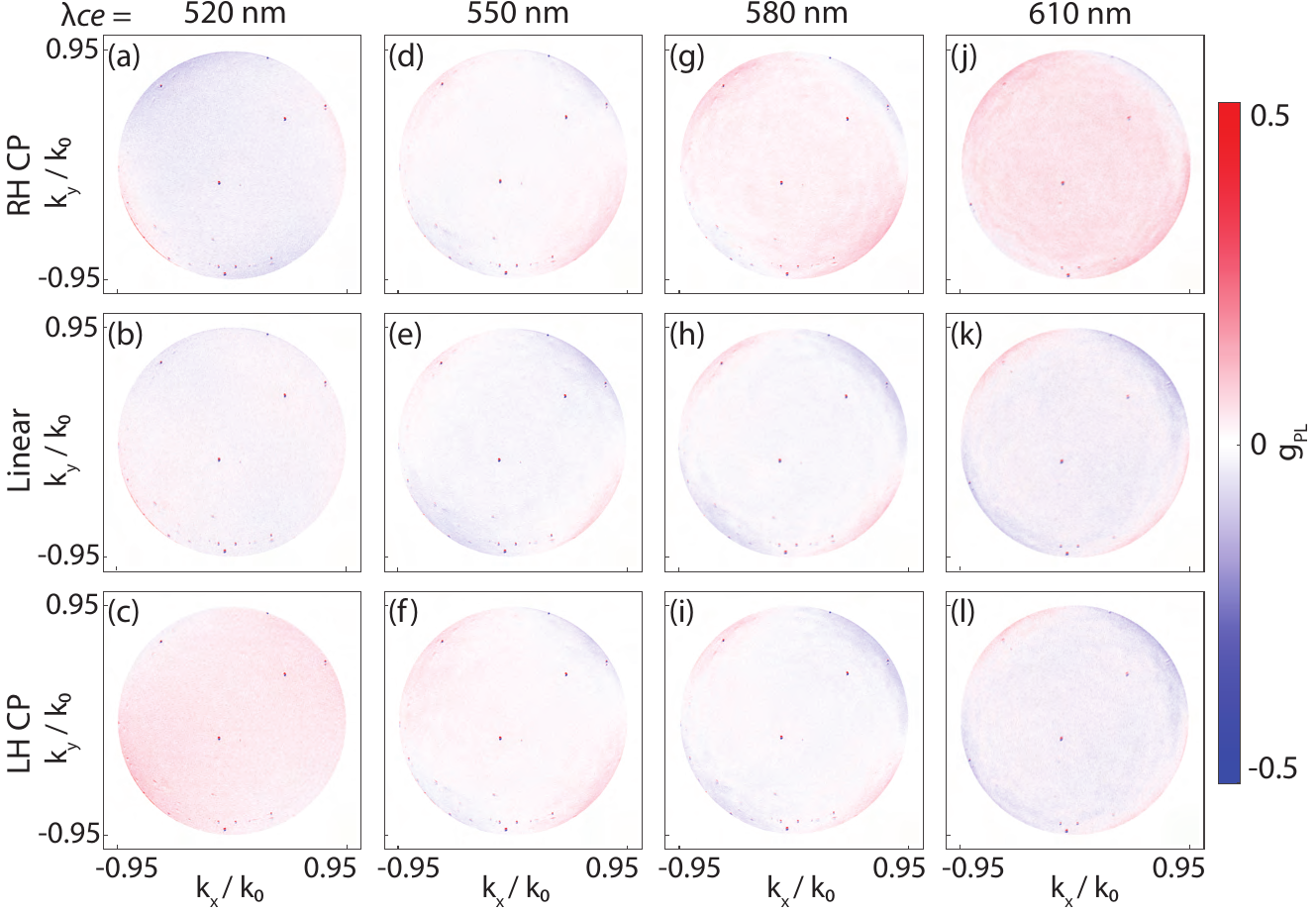}
  \caption{\textbf{Wavelength- and angle-resolved $\gPL$ maps of annealed F8BT:aza[6]H-($rac$) films.} Fourier-space $\gPL$ maps measured using 10-nm-FWHM band-pass filters centered at (a--c) 520~nm, (d--f) 550~nm, (g--i) 580~nm, and (j--l) 610~nm. Rows correspond to RH CP, linear, and LH CP excitation from top to bottom, respectively.}
\end{figure*}

\vspace{7cm}

Figs.~S4--S6 present the complete wavelength- and pump-polarization-resolved data sets for the annealed aza[6]H-($rac$), aza[6]H-($P$), and aza[6]H-($M$) films, respectively. The racemic blend remains nearly achiral across the measured wavelengths, whereas the enantiopure films exhibit opposite $\gPL$ signs and a pronounced dependence on both wavelength and observation angle, together with a modest pump-polarization memory effect.

\begin{figure*}[ht]
  \centering
  \includegraphics[width=0.9\textwidth]{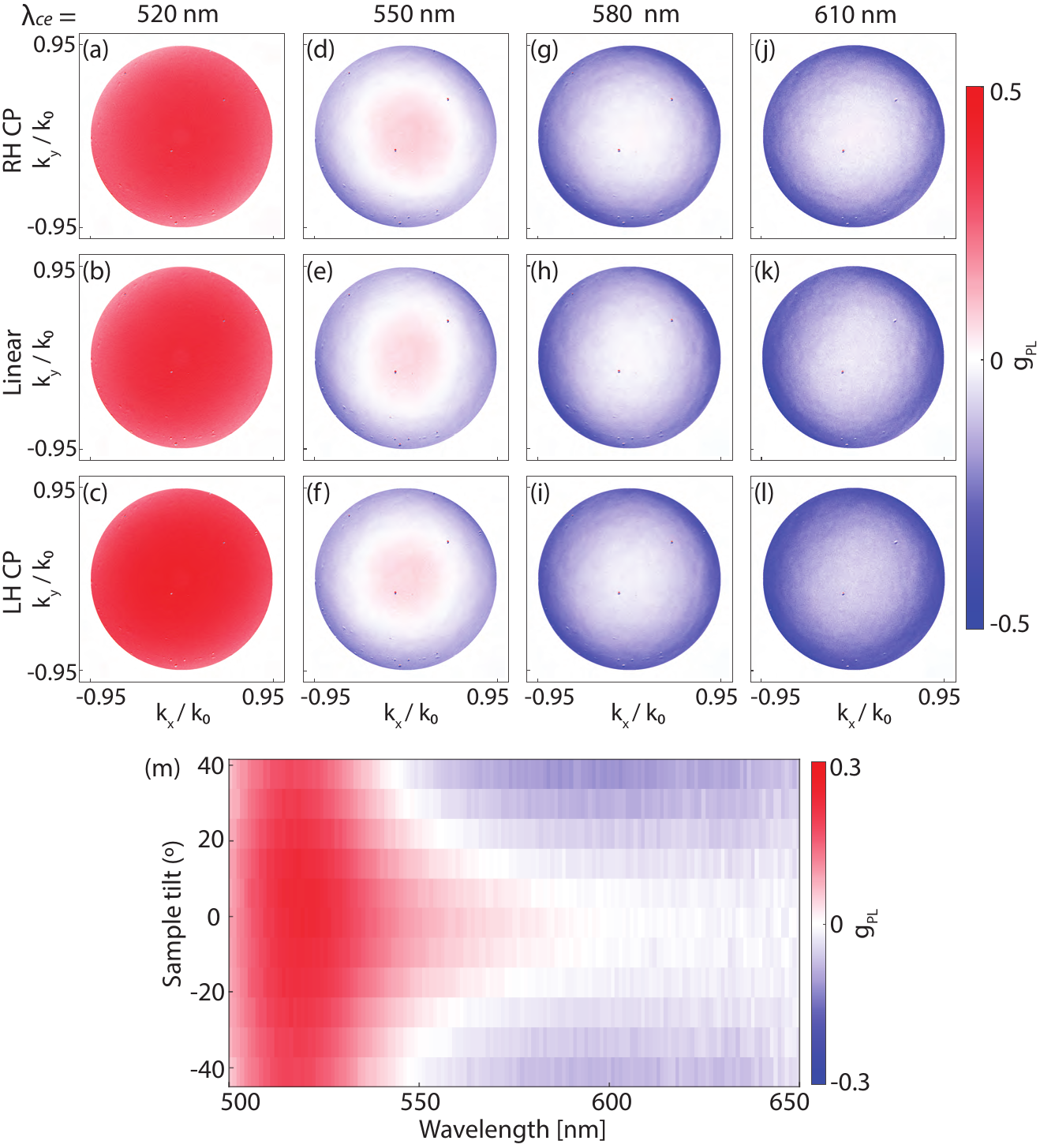}
  \caption{\textbf{Wavelength- and angle-resolved $\gPL$ maps of annealed F8BT:aza[6]H-($P$) films.} Fourier-space $\gPL$ maps measured using 10-nm-FWHM band-pass filters centered at (a--c) 520~nm, (d--f) 550~nm, (g--i) 580~nm, and (j--l) 610~nm. Rows correspond to RH CP, linear, and LH CP excitation from top to bottom, respectively. (m) Angle-dependent $\gPL$ measured using a conventional CPL spectrometer, plotted as a function of emission wavelength and sample tilt angle relative to the surface normal.}
\end{figure*}

\begin{figure*}[ht]
  \centering
  \includegraphics[width=0.9\textwidth]{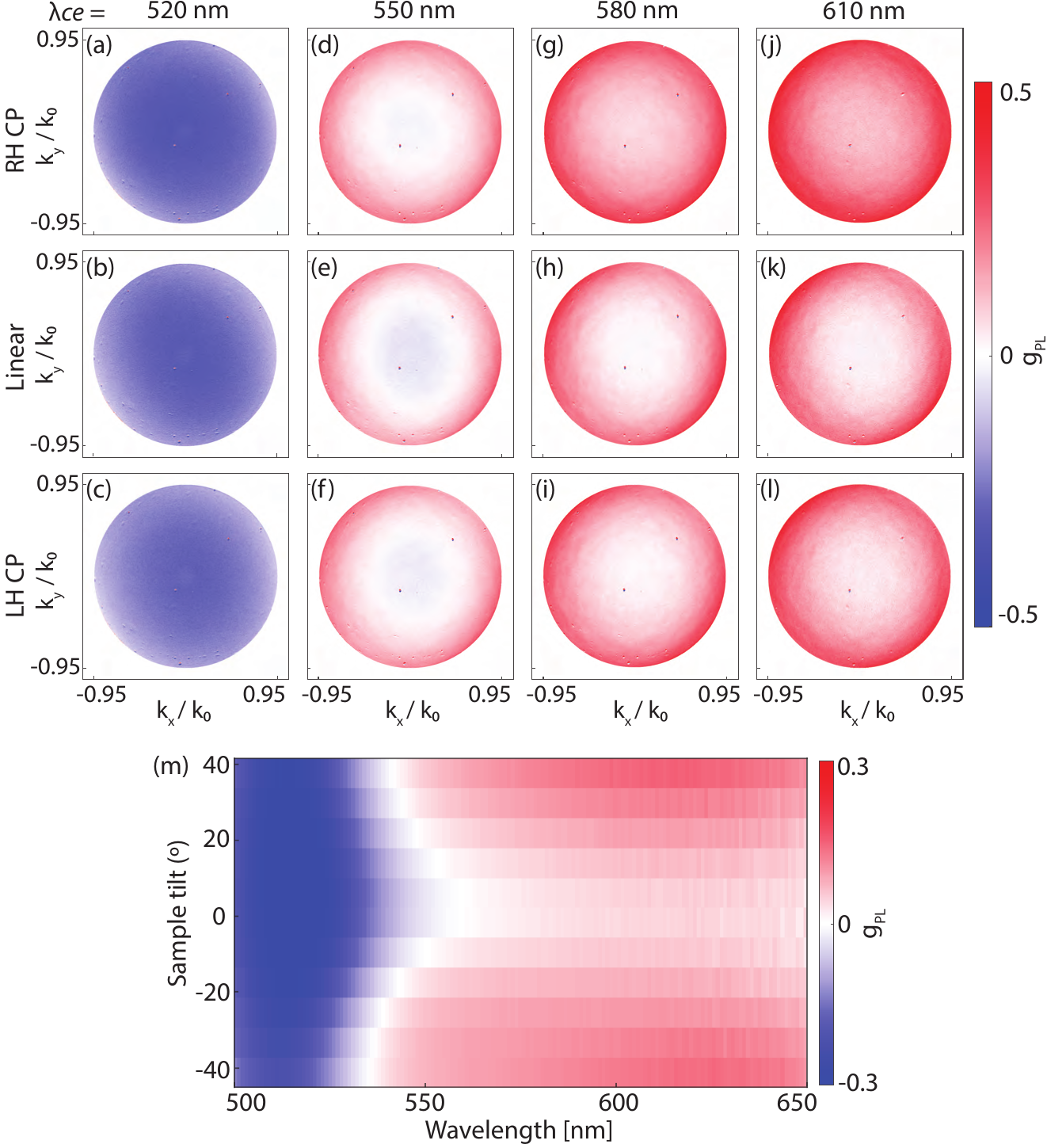}
  \caption{\textbf{Wavelength- and angle-resolved $\gPL$ maps of annealed F8BT:aza[6]H-($M$) films.} Fourier-space $\gPL$ maps measured using 10-nm-FWHM band-pass filters centered at (a--c) 520~nm, (d--f) 550~nm, (g--i) 580~nm, and (j--l) 610~nm. Rows correspond to RH CP, linear, and LH CP excitation from top to bottom, respectively. (m) Angle-dependent $\gPL$ measured using a conventional CPL spectrometer, plotted as a function of emission wavelength and sample tilt angle relative to the surface normal.}
\end{figure*}

\begin{figure*}[ht]
  \centering
  \includegraphics[width=0.8\textwidth]{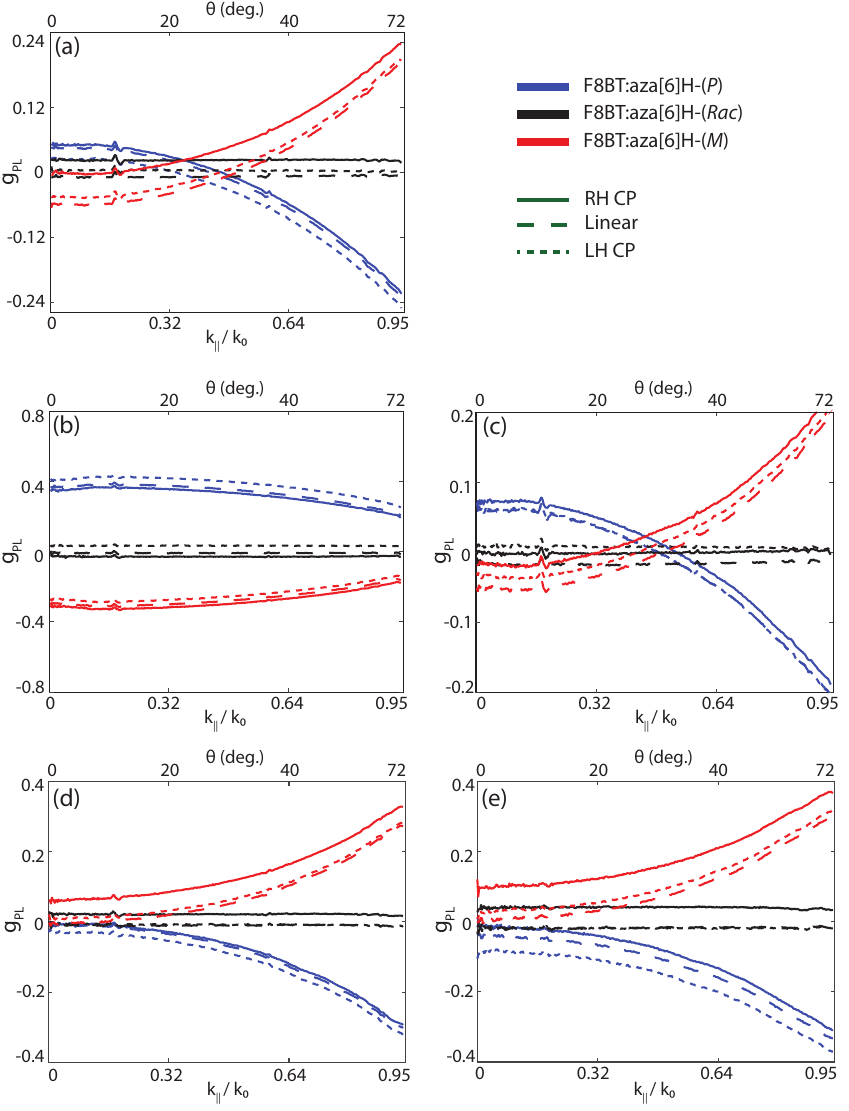}
  \caption{\textbf{Azimuthally averaged $\gPL$ from annealed F8BT:aza[6]H blend films.} $\gPL$ as a function of normalized in-plane momentum, $k_{\parallel}/k_0$ (bottom axis), and the corresponding observation angle (top axis) for films containing aza[6]H-($rac$) (red), aza[6]H-($P$) (green), and aza[6]H-($M$) (blue). (a) Panchromatic emission and (b--e) wavelength-resolved emission measured using 10-nm-FWHM band-pass filters centered at 520, 550, 580, and 610~nm, respectively. Solid, dashed, and dotted lines correspond to RH CP, linear, and LH CP excitation, respectively.}
\end{figure*}
\clearpage

To highlight the influence of pump polarization, the Fourier-space maps were azimuthally averaged as a function of normalized in-plane momentum. In Fig.~S7(a), the dashed curves correspond to linear excitation and reproduce the panchromatic angular profiles shown in Fig.~2f of the main text, while the solid and dotted curves show the corresponding response under RH CP and LH CP excitation, respectively. The resulting profiles show that pump handedness produces modest but systematic changes in the magnitude of $\gPL$, while preserving the dominant wavelength-, angle-, and enantiomer-dependent trends. For example, at 610~nm (Fig.~S6(e)), LH CP excitation enhances the predominantly RH CP emission from the F8BT:aza[6]H-($P$) film relative to RH CP excitation, whereas RH CP excitation enhances the predominantly LH CP emission from the F8BT:aza[6]H-($M$) film relative to LH CP excitation. At 520~nm (Fig.~S7(b)), where the intrinsic emission handedness is reversed, the F8BT:aza[6]H-($P$) film instead exhibits predominantly LH CP emission, enhanced under LH CP excitation, while the F8BT:aza[6]H-($M$) film exhibits predominantly RH CP emission, enhanced under RH CP excitation. Thus, the pump-polarization dependence is itself wavelength dependent and follows the spectral reversal of the chiral emission.

Polarization memory effects are well known for systems with linearly polarized emission dipoles (e.g., some organic dyes) where the excitation efficacy depends on the alignment between the absorption dipole and the pump polarization field, while the emission then follows the orientation of the transition dipole moment. This may give rise to emission with a preferred polarization along the pump polarization axis, depending on the relative orientation between the absorption and emission dipole moments --- provided that the molecular orientation remains unchanged during the fluorescence lifetime. In liquid environments, this effect is influenced and diminished due to rapid rotational diffusion times, whereas in the solid state, it persists (e.g., when molecular dipoles are immobilized within a polymer matrix). We interpret the observed dependence on pump polarization as an analogous effect for CP transition dipoles, wherein the handedness of excitation influences the emitted handedness through fixed chiral dipole orientations.

\clearpage
\section{Dipolar modeling of chiral emission in a stratified medium}
Intrinsic molecular chirality is often described by the dissymmetry factor ($g$-factor), given by~\cite{berova2000circular, han2018recent}:
\begin{equation}
g = 4\frac{\mathrm{Im}\!\left\{\mathbf{p}\cdot\mathbf{m}^{*}\right\}}
{|\mathbf{p}|^2+|\mathbf{m}|^2},
\end{equation}
where $\mathbf{p}$ and $\mathbf{m}$ are the electric and magnetic transition dipole moments, respectively. Here, $\mathbf{p}$ and $\mathbf{m}$ are considered in Gaussian (CGS) units; in SI units, the magnetic dipole moment must be rescaled by the speed of light $c$, i.e., $\mathbf{m}/c$, for unit consistency. 

The layered dielectric environment of the thin-film polymer blend --- with high refractive index ($n \approx 2$~\cite{wade2020natural}) sandwiched between air ($n = 1$) and glass ($n = 1.47$) --- leads to the preferential coupling of emission to parallel wavevectors ($k_{\parallel}$) beyond the total internal reflection angle, into guided modes of the polymer film. This results in a suppressed outcoupling fraction and possible angular polarization modulation due to unequal $s$- and $p$-polarized Fresnel coefficients.
\begin{figure*}[htbp!]
  \centering
  \includegraphics[width=0.82\textwidth]{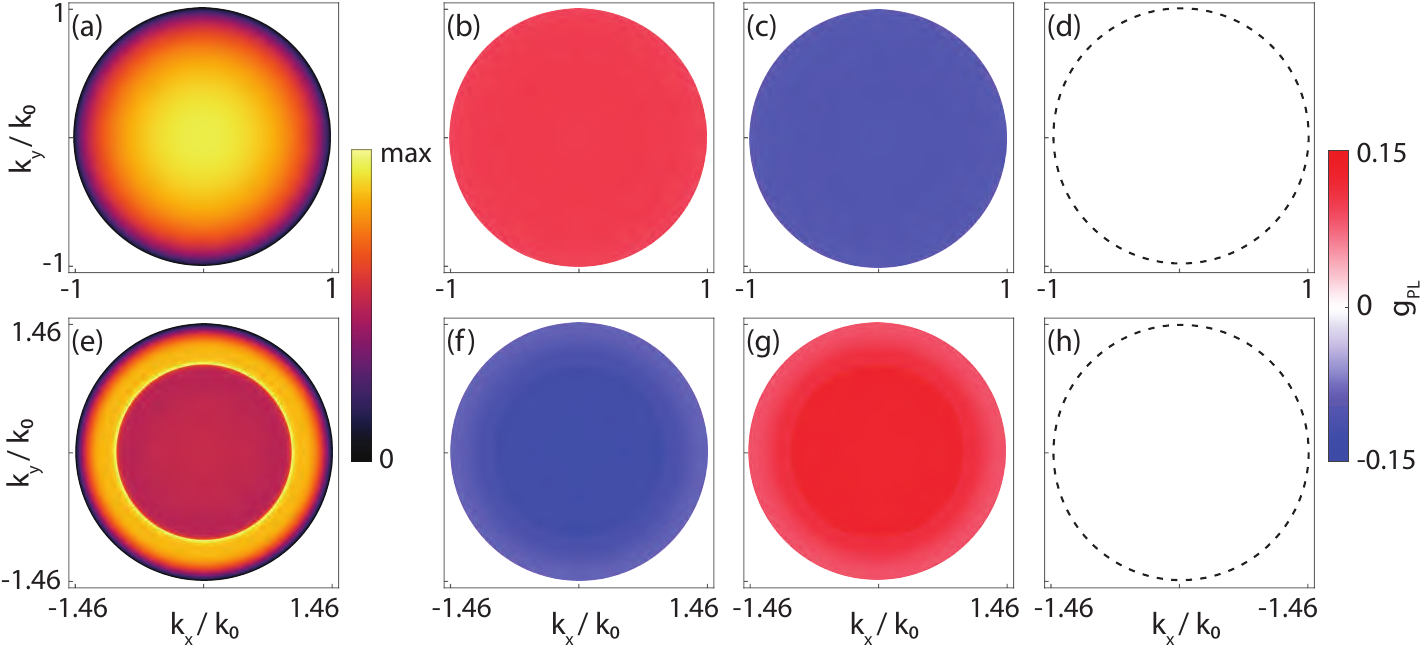}
  \caption{\textbf{Simulated angle-resolved $\gPL$ at $\lambda=600$~nm for an isotropic distribution of chiral dipoles.} The top and bottom rows correspond to emission into air and glass, respectively. (a,e) Total intensity; (b,f) $\gPL$ maps for handedness $\sigma=+1$; (c,g) $\sigma=-1$; (d,h) achiral case $\sigma=0$. The apparent handedness reverses between the two hemispheres due to the change in viewing direction.}
  \label{fig:PlanarSimulation}
\end{figure*}
To isolate these effects, we perform simulations using the Green's function formalism in a multilayer system, accounting for the reactive back-action of reflected fields via $k$-dependent Fresnel coefficients. Emission is modeled within our in-house Python-based PyRAMIDS package~\cite{pyramids}, assuming an isotropic distribution of dipole orientations located at the center of a 175~nm-thick film ($n = 2$). We calculate far-field radiation patterns and angular $\gPL$ maps by incoherently averaging over all dipole alignments.

In archetypical models of chirality, the electric and magnetic dipole moments are assumed to be parallel and exhibit a quarter-wave phase difference ($\pi/2$ out of phase), with handedness encoded by the phase sign, while the relative magnitude determines the strength of chirality. We use $(\mathbf{p}, \mathbf{m}) = (1, 0.04 \sigma i)$, with $\sigma = \pm 1$ denoting opposite handedness and $\sigma = 0$ for the achiral case. The magnitude of $\mathbf{m}$ was chosen to reproduce the order of magnitude of the experimentally measured emission dissymmetry. The simulated total emission intensity is approximately Lambertian in air (Fig.~\ref{fig:PlanarSimulation}a) and exhibits a bright emission ring in the glass substrate at $k_\parallel/k_0>1$, beyond the air light line (Fig.~~\ref{fig:PlanarSimulation}e). For the $g$-parameter maps, we observe handedness reversal with $\sigma$ (Figs.~\ref{fig:PlanarSimulation}b–c, f–g), while the achiral case yields no chiral emission (Figs.~\ref{fig:PlanarSimulation}d, h). The apparent reversal of handedness between air-side and glass-side emission arises from the reversal of the observation direction relative to the emitter plane, rather than from a physical conversion of the emitted helicity. Interestingly, while the simulations reproduce the expected direction-dependent flip in handedness between hemispheres, they do not predict strong angular variation in $\gPL$. This suggests that thin-film interference and dielectric contrast alone cannot fully account for the experimentally observed angular modulation, pointing toward additional chiral photonic or supramolecular effects.

\clearpage
\section{Finite-element simulations of metasurface-coupled emission}
\begin{figure*}[ht]
  \centering
  \includegraphics[width=0.75\textwidth]{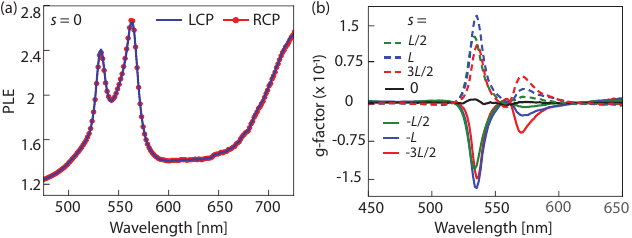}
  \caption{\textbf{COMSOL simulation of the metasurface optical response under normal incidence.} (a) Predicted PLE for an achiral polymer film on an achiral metasurface ($s=0$). (b) Predicted $\gPL$ spectra for metasurfaces with different lateral shifts $s$.}
  \label{fig:COMSOL}
\end{figure*}
We implement the Helmholtz reciprocity principle via finite-element simulations in COMSOL Multiphysics v5.2.0 to estimate polarization-resolved PLE from an achiral polymer embedded with a metasurface. The simulation domain consists of a single periodic unit cell with Bloch–Floquet boundary conditions applied laterally and perfectly matched layers (PMLs) placed at the excitation and collection ports in the vertical direction. Nanostructures are meshed at 5~nm resolution, while other layers are discretized at one-sixth of the local wavelength. According to Lorentz reciprocity, the emission from a dipole distribution into a given far-field direction and polarization can be related to the local fields generated in the emitter layer by illumination from the reciprocal direction and polarization. To estimate far-field PLE, we introduce a weak tracer loss ($k = 0.01$) in the polymer layer and compute the volume-integrated absorption for different input polarizations, normalizing against the absorption in a unpatterned film --- without a metasurface. For an achiral unit cell ($s = 0$), the PLE is identical for LH CP and RH CP excitation, yielding a negligible $\gPL$. The polymer layer supports two waveguide modes with transverse magnetic (TM) and transverse electric (TE) polarizations. Two distinct resonant features near 535~nm and 560~nm are associated with TM- and TE-like waveguide modes, respectively, while a broader LSPR appears at $\lambda>700$~nm. Introducing planar chirality ($s \ne 0$) breaks in-plane mirror symmetry and produces non-zero $\gPL$ values reaching up to 0.15. Notably, the sign of $\gPL$ reverses with the sign of the rod shift, demonstrating geometric control over emission handedness. Although the $s\neq0$ structures exhibit planar chiroptical signatures, they remain invariant under out-of-plane reflection followed by a $180^\circ$ rotation and are therefore not truly chiral in three dimensional (3D) space. Consequently, the chiroptical asymmetry reverses sign between the lower and upper hemispheres, corresponding to observation from the glass-substrate and air-superstrate sides, respectively.

\section{Extended angle-resolved measurements of metasurface-coupled films}
Figs.~S10--S12 show the complete pump-polarization-dependent $\gPL$ data sets for the racemic and enantiopure blend films coupled to metasurfaces with different lateral shifts $s$. These measurements complement Fig.~4 of the main text and show that the metasurface-induced angular dissymmetry is robust across the three pump polarization states. Reversing the sign of the unit-cell lateral shift, $s$, reverses the dominant metasurface-induced emission handedness, while the enantiopure films retain an additional contribution from the intrinsic chirality of the blend.

\,\,
\begin{figure*}[ht]
  \centering
  \includegraphics[width=\textwidth]{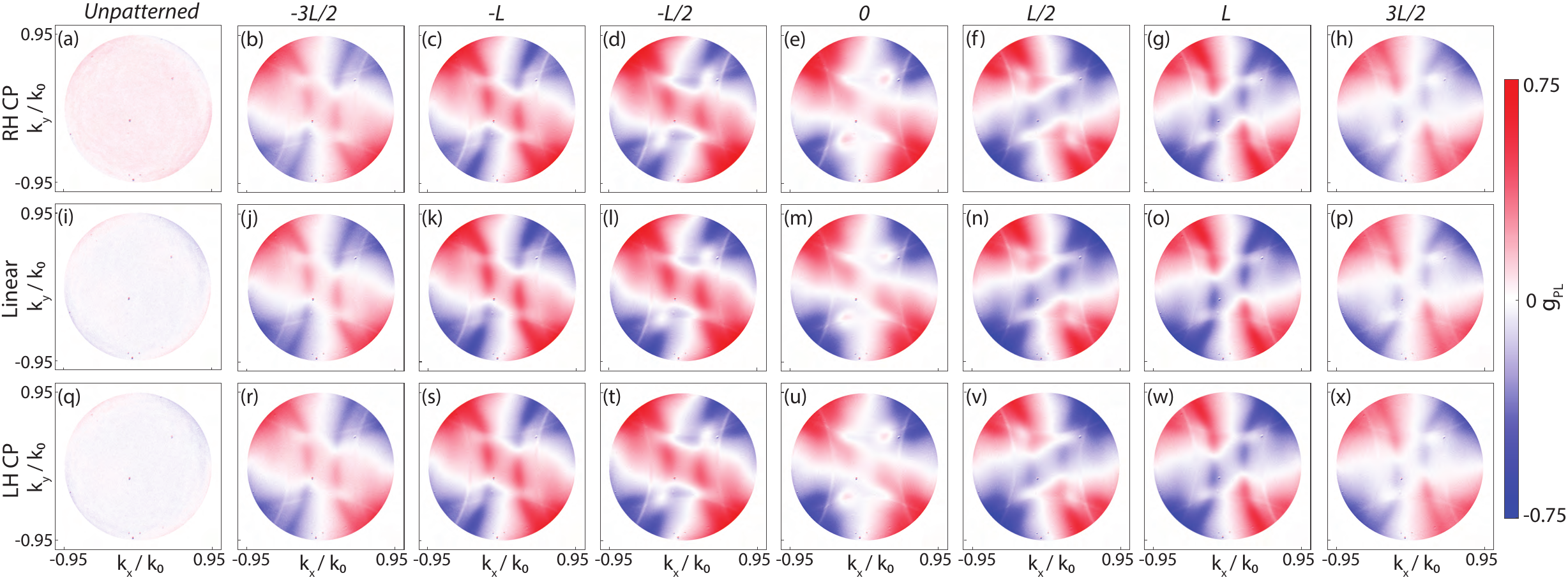}
  \caption{\textbf{Angle-resolved $\gPL$ modification in F8BT:aza[6]H-($rac$) films induced by metasurface coupling.} Fourier-space $\gPL$ maps at $\lambda = 610$~nm measured under different pump polarizations (rows: RH CP, linear, and LH CP). Columns correspond to the unpatterned-film reference and to metasurfaces with lateral shifts $s = -3L/2$, $-L$, $-L/2$, $0$, $L/2$, $L$, and $3L/2$, respectively.}
\end{figure*}

\begin{figure*}[ht]
  \centering
  \includegraphics[width=\textwidth]{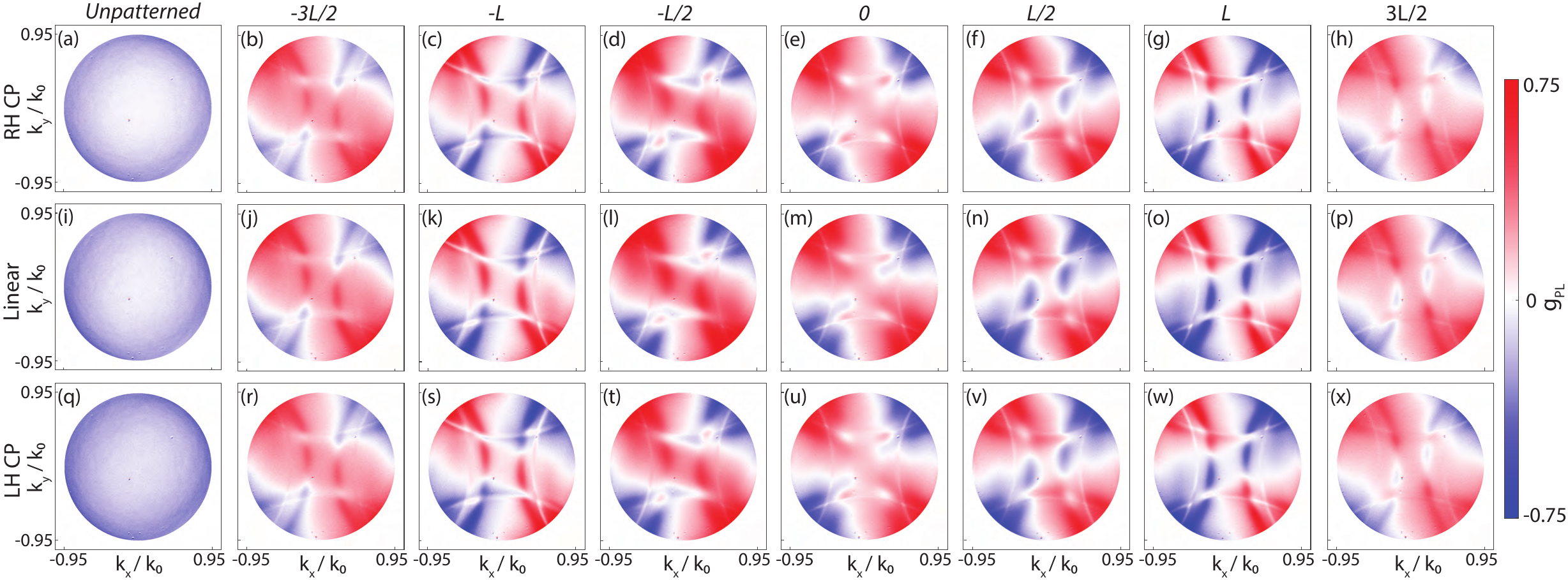}
  \caption{\textbf{Angle-resolved $\gPL$ modification in F8BT:aza[6]H-($P$) films induced by metasurface coupling.} Fourier-space $\gPL$ maps at $\lambda = 610$~nm measured under different pump polarizations (rows: RH CP, linear, and LH CP). Columns correspond to the unpatterned-film reference and to metasurfaces with lateral shifts $s = -3L/2$, $-L$, $-L/2$, $0$, $L/2$, $L$, and $3L/2$, respectively.}
\end{figure*}

\begin{figure*}[ht]
  \centering
  \includegraphics[width=\textwidth]{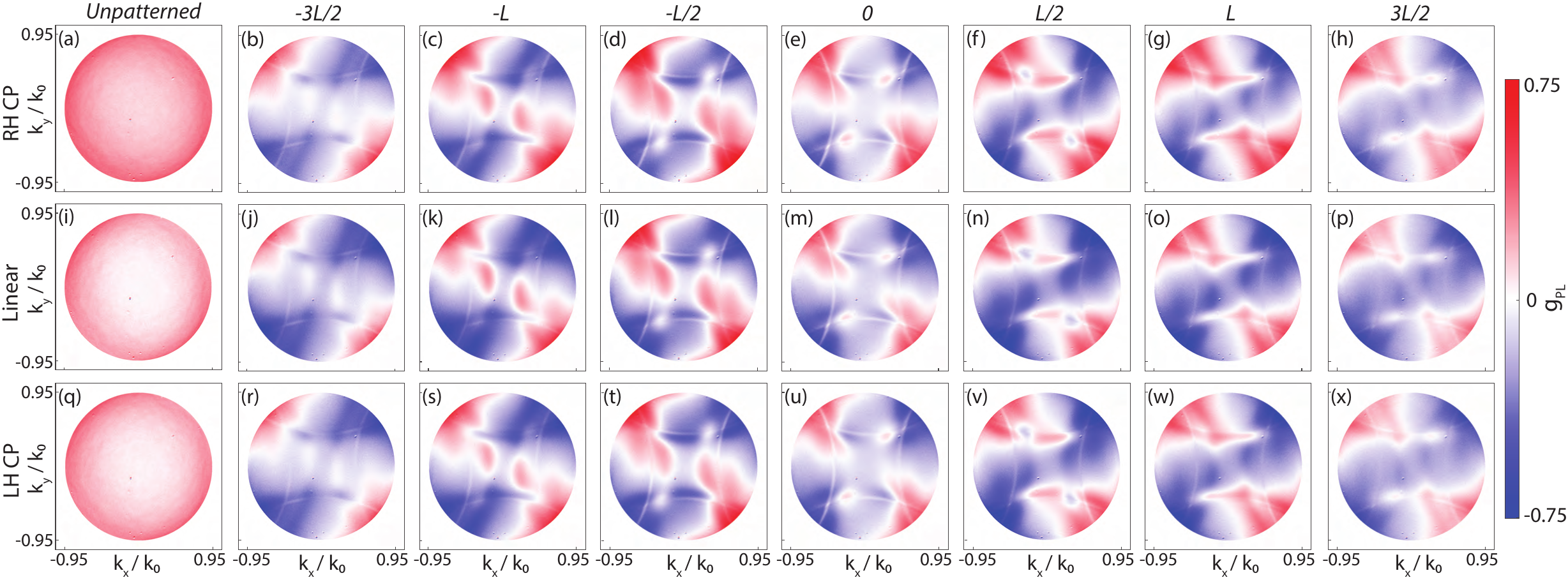}
  \caption{\textbf{Angle-resolved $\gPL$ modification in F8BT:aza[6]H-($M$) films induced by metasurface coupling.} Fourier-space $\gPL$ maps at $\lambda = 610$~nm measured under different pump polarizations (rows: RH CP, linear, and LH CP). Columns correspond to the unpatterned-film reference and to metasurfaces with lateral shifts $s = -3L/2$, $-L$, $-L/2$, $0$, $L/2$, $L$, and $3L/2$, respectively.}
\end{figure*}

\clearpage
\section{Experimental setup for angle-resolved emission measurements}
\begin{figure*}[ht]
  \centering
  \includegraphics[width=0.67\textwidth]{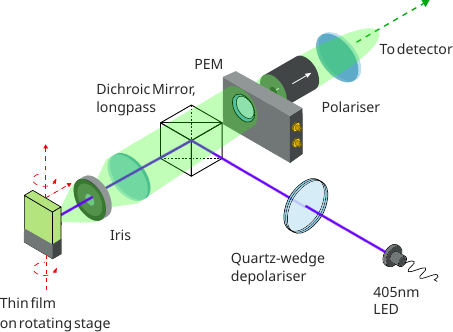}
  \caption{\textbf{Schematic of the custom-built CPL spectrometer.}
A depolarized 405~nm LED excites the sample mounted on a rotating stage, and the emitted light is analyzed using a photoelastic modulator (PEM) and polarizer before being directed to the spectrometer and detector.}
\end{figure*}  
Unpatterned films on glass were excited using light from a 405 nm LED passed through a quartz-wedge depolarizer,  using a dichroic mirror  (Thorlabs DMLP425) to direct light onto the sample from the front. Luminescence from the sample was collimated by a 2-inch diameter lens and directed through a PEM (PEM-100, Hinds Instruments) set to quarter-wave retardation, followed by a Glan-Thompson polarizer (Thorlabs) at 45$^{o}$ to the PEM axis. An iris was used to block any light that would pass around the edges of the polarizer. A second lens focused the luminescence into the input slit of a monochromator (Bentham TMC-300). Wavelength-selected light was then detected using a photomultiplier tube (PMT, Hamamatsu H13226A-103), which produced a current signal proportional to light intensity. This was converted to a voltage signal using a low-noise transimpedance amplifier (Femto), the signal from which was passed to a lock-in amplifier (Signaloc 2100, Hinds Instruments). The sample was mounted at the focal point of the first lens on a rotatable mount, which allowed the manual adjustment of angle using a micrometer scale. The instrument was controlled using custom Python software and obtained $\gPL$ as described in~\cite{moreno2024enhancing}. 

\begin{figure*}[ht]
  \centering
  \includegraphics[width=0.65\textwidth]{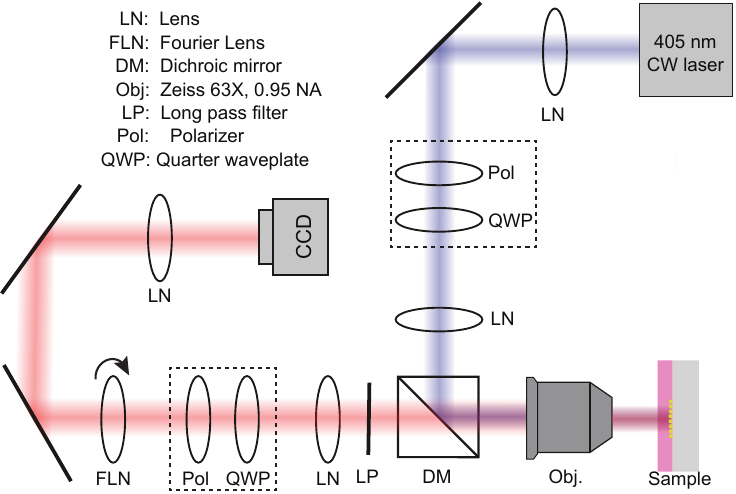}
  \caption{\textbf{Schematic of the Fourier polarimetry setup.} A 405~nm CW laser excites the sample through a high-NA objective. Polarimeters in the excitation and detection paths control the pump polarization and analyze the polarization-resolved emission, respectively.}
\end{figure*}  
Fourier microscopy measurements were performed using a custom-built inverted microscope~\cite{pal2025dark}, with both pump excitation and detection through the high NA objective (Zeiss N-Achroplan 63X, 0.95~NA, no coverglass) from the air/polymer interface side. A 405~nm continuous wave (CW) laser (Cobolt 06-MLD) was directed to excite the sample after reflecting from a dichroic mirror (Chroma, ZT405rdc-UF1-25x36), with an epi lens in the pump path producing a $\sim75~\mu$m beam diameter on the sample plane; the incident pump power was measured as 2 $\mu$W at the sample plane using a photodiode placed after the objective (Thorlabs S121C). Fluorescence was filtered using a 425~nm longpass filter (Chroma ET425lp) and imaged onto a thermoelectrically cooled CCD camera (Andor Clara) via a 200~mm tube lens and telescope. To acquire angular (Fourier-space) emission patterns, the second telescope lens was removed so that the first lens, placed one focal length from the objective's back focal plane, served as a Fourier lens. Polarization control of the pump light was achieved using a polarizer (WP25M-VIS) and a quarter-wave plate (AQWP05M-600) in the excitation path. An identical polarimeter was placed in the detection path to analyze fluorescence polarization. The collected fluorescence may undergo slight polarization changes due to the dichroic beam splitter and reflections from the silver-coated mirrors (PF10-03-P01). The circular-emission dissymmetry was obtained from the reconstructed Stokes parameters as $\gPL = 2S_3/S_0$, with the sign convention chosen such that positive $\gPL$ corresponds to predominantly LH CP emission. The Fourier-polarimetry methodology follows the approach reported previously in Ref.~\cite{pal2025dark}.

\bibliography{references.bib}